\documentclass[aps, prd, amsmath, floats, floatfix, onecolumn, nofootinbib, nosuperscriptaddress]{revtex4-2}
\usepackage[pdftex]{graphicx}
\usepackage{color}
\usepackage{latexsym}
\usepackage[colorlinks]{hyperref}
\usepackage{amsthm,amsmath,amssymb}
\usepackage{mathrsfs}
\usepackage{makecell}

\newcommand{\be}{\begin{eqnarray}}
\newcommand{\ee}{\end{eqnarray}}
\newcommand{\rar}{\rightarrow}

\begin{document}

\title{Black hole X-ray spectra: notes on the relativistic calculations}

\author{Cosimo~Bambi}
\email[]{bambi@fudan.edu.cn}
\affiliation{Center for Astronomy and Astrophysics, Center for Field Theory and Particle Physics, and Department of Physics,\\
Fudan University, Shanghai 200438, China}
\affiliation{School of Natural Sciences and Humanities, New Uzbekistan University, Tashkent 100007, Uzbekistan}

\begin{abstract}
This is a collection of notes to calculate electromagnetic spectra of geometrically thin and optically thick accretion disks around black holes. The presentation is intentionally pedagogical and most calculations are reported step by step. In the disk-corona model, the spectrum of a source has three components: a thermal component from the disk, a Comptonized component from the corona, and a reflection component from the disk. These notes review only the relativistic calculations. The formulas presented here are valid for stationary, axisymmetric, asymptotically-flat, circular spacetimes, so they can be potentially used for a large class of black hole solutions. 
\end{abstract}

\maketitle

%%%%%%%%%%%%%%%%%%%%%%%%%%%%%%

\section{Introduction}

The development of sufficiently advanced astrophysical models is crucial if we want to analyze high-quality data. The aim of these notes is to review the relativistic calculations of the electromagnetic spectrum of thin accretion disks around black holes.

\begin{figure}[b]
\centering
\includegraphics[width=0.45\linewidth]{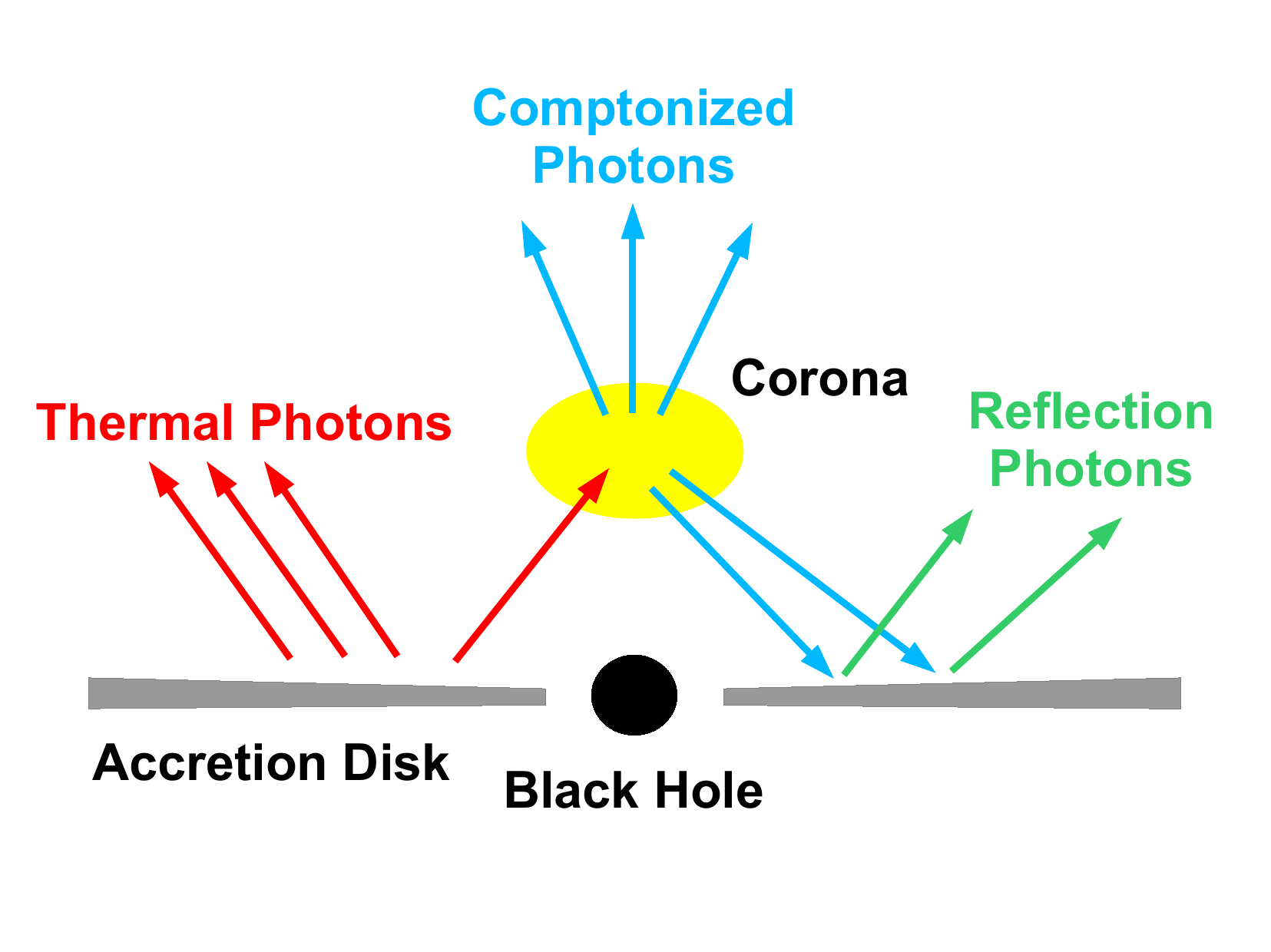}
\hspace{0.5cm}
\includegraphics[width=0.45\linewidth]{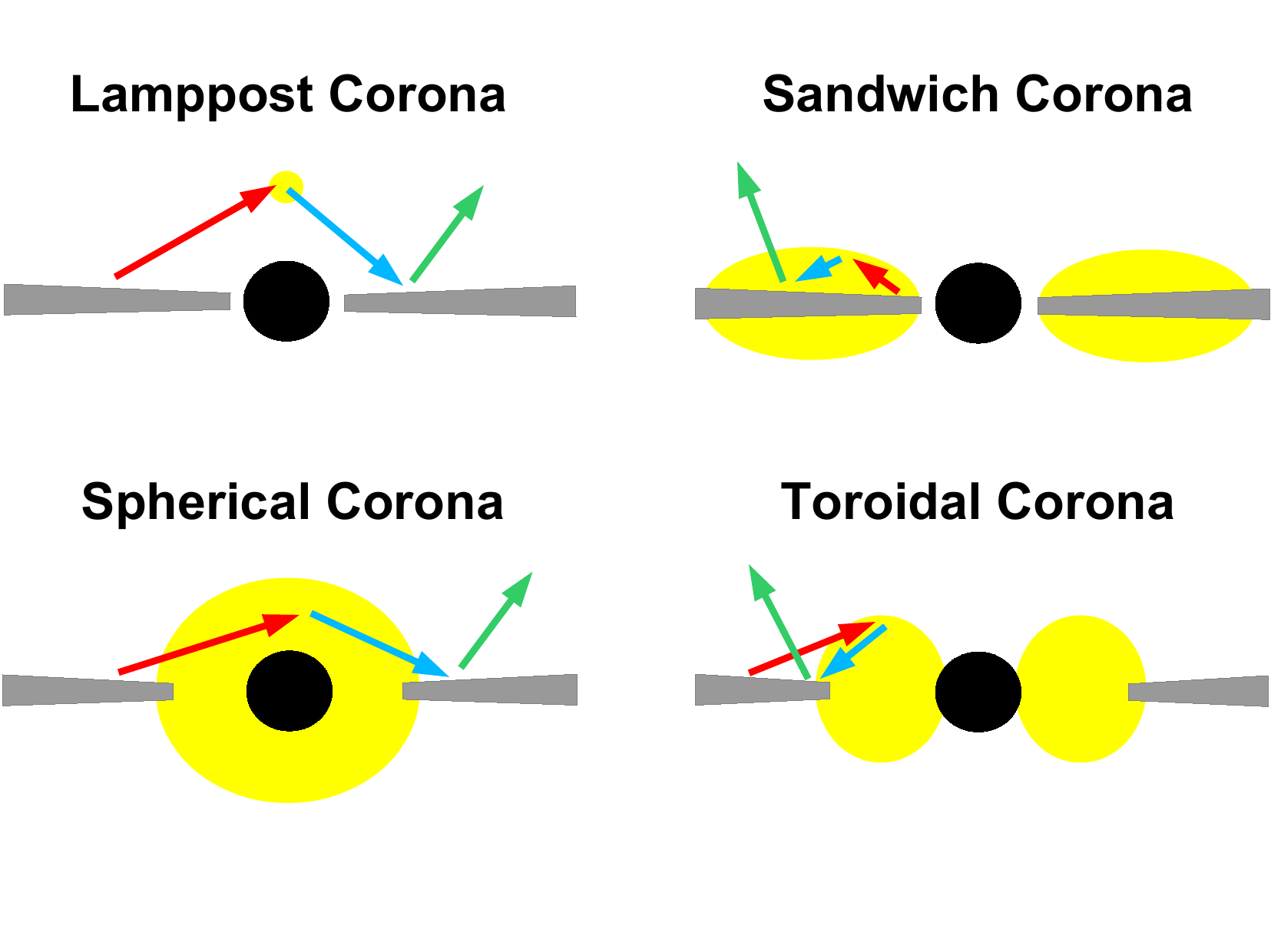}
\vspace{-1.0cm}
\caption{Left panel: Disk-corona model. Right panel: Examples of possible coronal geometries. See the text for more details.}
\label{f-corona}
\end{figure}

The physical system that we want to study is shown in the left panel in Fig.~\ref{f-corona} and is normally referred to as the {\it disk-corona model}. The black hole can be either a stellar-mass black hole in an X-ray binary or a supermassive black hole in an active galactic nucleus. The key-point is that the black hole accretes from a cold, geometrically thin, and optically thick accretion disk\footnote{Generally speaking, an accretion disk is geometrically thin if $h/r \ll 1$, where $h$ is the thickness of the disk at the radial coordinate $r$, and is optically thick is $h \gg \lambda$, where $\lambda$ is the photon mean free path in the disk. The accretion disk is instead geometrically thick if $h/r \gtrsim 1$. The accretion disk is optically thin if $h \ll \lambda$.}. The disk is ``cold'' because it can efficiently emits radiation. Every point on the surface of the accretion disk has a blackbody-like spectrum and the whole accretion disk has a multi-temperature blackbody-like spectrum. The thermal spectrum of the disk is normally peaked in the soft X-ray band (0.1-10~keV) for stellar-mass black holes in X-ray binaries and in the UV band (1-100~eV) for supermassive black holes in active galactic nuclei~\cite{Shakura:1972te,Page:1974he}. The {\it corona} (yellow regions in Fig.~\ref{f-corona}) is some ``hot'' ($\sim 100$~keV) electron cloud close to the black hole and the inner part of the accretion disk, but its exact morphology is not yet well understood~\cite{Bisnovatyi-Kogan:1977xxx,Haardt:1991tp,Dove:1997ei,Liu:2003yg,Markoff:2005ht,Sironi:2019sxv}. The right panel in Fig.~\ref{f-corona} shows possible coronal geometries. In the lamppost model, the corona is a compact source along the black hole spin axis (for example, the base of the jet may act as a lamppost corona~\cite{Markoff:2005ht}). In the sandwich model, the corona is the hot atmosphere above the accretion disk. In the spherical and toroidal models, the corona is the hot material in the plunging region, between the inner edge of the accretion disk and the black hole. The coronal geometry can change (even on timescales of order of hours or days in the case of X-ray binaries) and two or more coronae may coexist at the same time.

Since the disk is cold and the corona is hot, thermal photons from the disk (red arrows in Fig.~\ref{f-corona}) can inverse Compton scatter off free electrons in the corona. The spectrum of the Comptonized photons can be normally approximated well by a power law with a low and a high energy cutoff~\cite{Zdziarski:1996wq,Zdziarski:2019cvs}. A fraction of the Comptonized photons (blue arrows in Fig.~\ref{f-corona}) can illuminate the disk: Compton scattering and absorption followed by fluorescent emissions produce the reflection spectrum (green arrows in Fig.~\ref{f-corona}).

The reflection spectrum in the rest-frame of the material in the disk is characterized by narrow fluorescent emission lines in the soft X-ray band and a Compton hump with a peak around 20-40~keV~\cite{Ross:2005dm,Garcia:2010iz}. The strongest emission feature is often the iron K$\alpha$ complex, which is a very narrow feature at 6.4~keV in the case of neutral or weakly ionized iron atoms and shifts up to 6.97~keV in the case of H-like iron ions. The reflection spectrum of the whole disk as seen far from the source is blurred because of relativistic effects (Doppler boosting due to the motion of the material in the disk and gravitational redshift due to the gravitational well of the black hole)~\cite{Fabian:1989ej,Laor:1991nc,Bambi:2017khi}. {\it X-ray reflection spectroscopy} is the analysis of these relativistically blurred reflection features in the X-ray spectra of X-ray binary systems and active galactic nuclei and can be a powerful tool to study the accretion process around black holes, measure black hole spins~\cite{Brenneman:2006hw,Bambi:2020jpe,Draghis:2023vzj}, and test Einstein's theory of General Relativity in the strong field regime~\cite{Cao:2017kdq,Tripathi:2018lhx,Tripathi:2020yts}.

In these notes, we assume that the spacetime is stationary, axisymmetric, asymptotically-flat, and circular. In spherical-like coordinates $(t,r,\theta,\phi)$, the line element can always be written in the following form
\be\label{eq-metric}
ds^2 = g_{tt} dt^2 + 2 g_{t\phi} dt \, d\phi + g_{rr} dr^2 + g_{\theta\theta} d\theta^2 + g_{\phi\phi} d\phi^2 \, ,
\ee 
where the metric coefficients are independent of the coordinates $t$ and $\phi$. The fact that the only non-vanishing off-diagonal element is $g_{t\phi}$ is a consequence of the assumption that the spacetime is {\it circular}~\cite{Papapetrou:1966zz,Wald:1984rg}. Without such an assumption, in general we may have other non-vanishing off-diagonal metric coefficients. In General Relativity, $R_{\mu\nu} = 0$ in vacuum and we can always write the line element as in Eq.~(\ref{eq-metric}) for vacuum solutions if the spacetime is stationary and axisymmetric~\cite{Wald:1984rg}.

In these notes, we employ units in which $G_{\rm N} = c = \hbar = 1$, unless stated otherwise, and the convention of a metric with signature $(-+++)$.

%%%%%%%%%%%%%%%

\subsection{Basic concepts}\label{ss-basic}

\begin{figure}[b]
\centering
\includegraphics[width=0.45\linewidth]{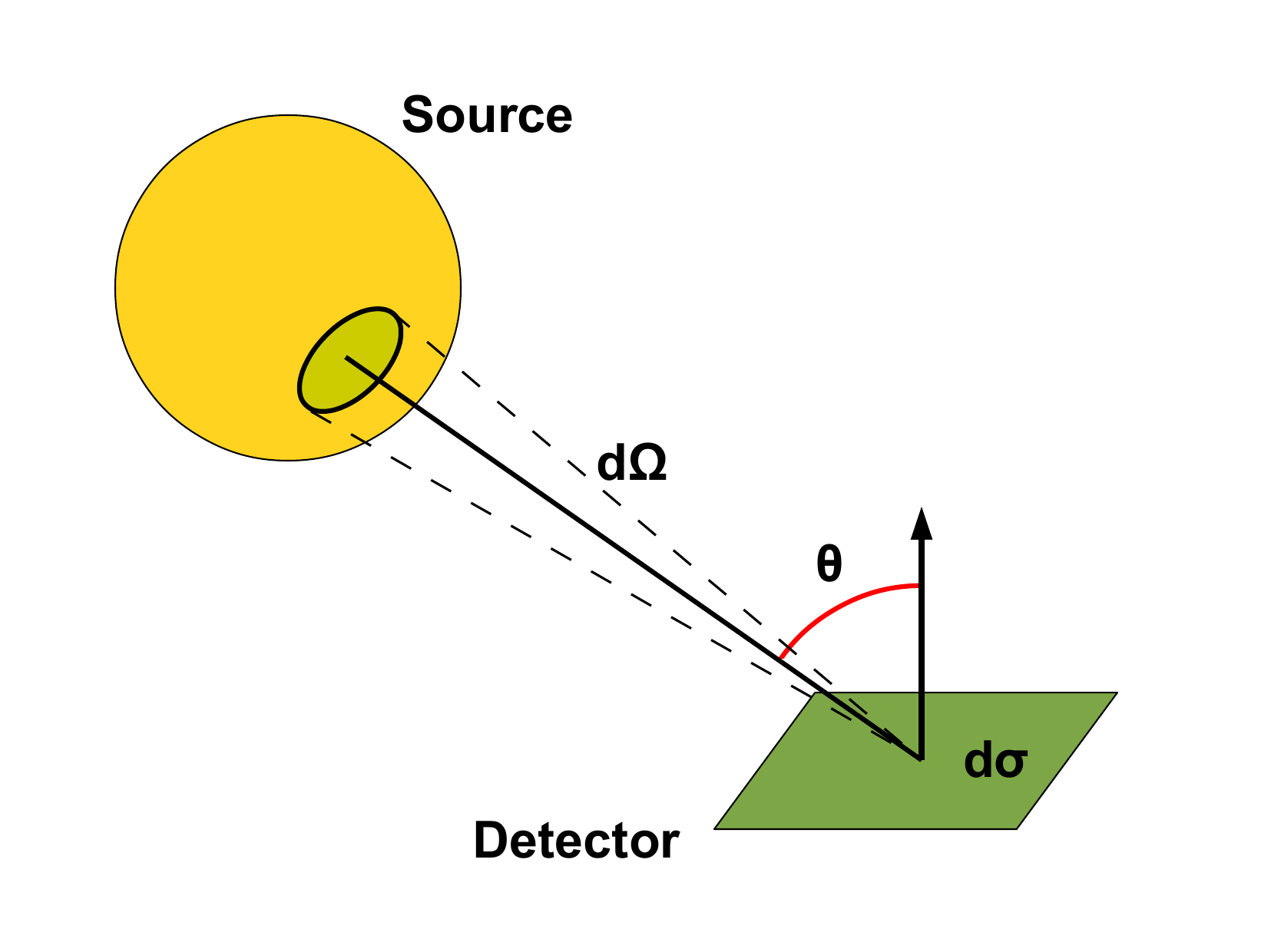}
\hspace{-0.5cm}
\includegraphics[width=0.45\linewidth]{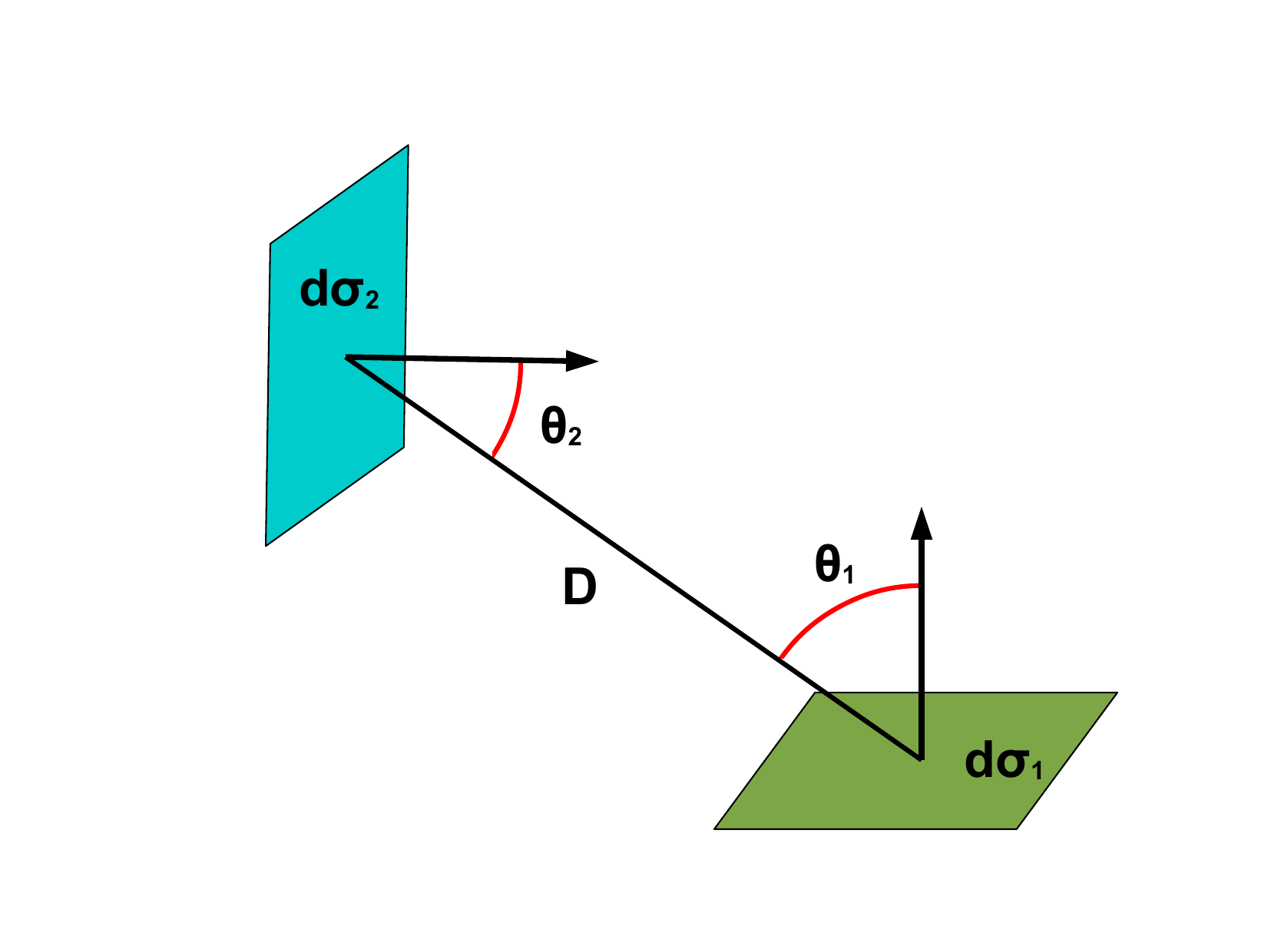}
\vspace{-0.5cm}
\caption{Left panel: The source illuminates the detector and deposits the energy $dE$ on the surface $d\sigma$ of the detector from the solid angle $d\Omega$ in the time $dt$ and in the frequency band of width $d\nu$: $dE = I_\nu \, \cos\theta \, d\sigma \, d\Omega \, dt \, d\nu$. Right panel: the specific intensity $I_\nu$ and the total intensity $I$ are conserved along photon trajectories in empty space. See the text for more details.}
\label{f-intensity}
\end{figure}

It is useful to start reviewing some basic concepts. For the moment, we ignore relativistic effects. Let us consider a detector and a source, as shown in the left panel in Fig.~\ref{f-intensity}. The detector and the source are in empty space, so there is no absorption, scattering, or emission of radiation along the path from the source to the detector. Moreover, the size of the detector, the size of the source, and the distance between the source and the detector are much larger than the wavelength of the radiation emitted by the source, so we can use the {\it ray-optics approximation}, where the radiated energy flows in straight lines. $\theta$ is the angle between the straight line connecting the detector and the source and the normal to the surface of the detector. The energy $dE$ illuminating the infinitesimal surface $d\sigma$ of the detector from the solid angle $d\Omega$ in the time $dt$ and in the frequency band of width $d\nu$ is
\be
dE = I_\nu \, \cos\theta \, d\sigma \, d\Omega \, dt \, d\nu \, ,
\ee
where we have introduced the {\it specific intensity} of the radiation $I_\nu$, which is thus defined as
\be
I_\nu = \frac{dE}{\left( \cos\theta \, d\sigma \right) d\Omega \, dt \, d\nu} \, .
\ee
The specific intensity can be measured, for example, in W~m$^{-2}$~Hz$^{-1}$~sr$^{-1}$. The {\it total intensity} is obtained after integrating the specific intensity over all frequencies
\be
I = \int_0^{\infty} I_\nu \, d\nu \, .
\ee
The total intensity can be measured, for example, in W~m$^{-2}$~sr$^{-1}$.

From our definition of specific intensity and total intensity, these two quantities are conserved along photon trajectories in empty space. Let us consider two infinitesimal surfaces, $d\sigma_1$ and $d\sigma_2$, as shown in the right panel in Fig.~\ref{f-intensity}. $D$ is the distance between $d\sigma_1$ and $d\sigma_2$. $d\Omega_1 = \cos\theta_2 d\sigma_2 / D^2$ is the infinitesimal solid angle subtended by $d\sigma_2$ as seen from the surface $d\sigma_1$ and $d\Omega_2 = \cos\theta_1 d\sigma_1 / D^2$ is the infinitesimal solid angle subtended by $d\sigma_1$ as seen from the surface $d\sigma_2$. The power $dE_1/dt$ in the frequency range $\nu$ to $\nu + d\nu$ flowing through the surface $d\sigma_1$ in the solid angle $d\Omega_1$ is
\be
\frac{dE_1}{dt} = I_\nu^1 \, \cos\theta_1 \, d\Omega_1 \, d\sigma_1 \,  d\nu =
I_\nu^1 \, \cos\theta_1 \, \frac{\cos\theta_2 \, d\sigma_2}{D^2} \, d\sigma_1 \,  d\nu \, ,
\ee
where $I_\nu^1$ is the specific intensity on $d\sigma_1$. In the same way, the power $dE_2/dt$ in the frequency range $\nu$ to $\nu + d\nu$ flowing through the surface $d\sigma_2$ in the solid angle $d\Omega_2$ is
\be
\frac{dE_2}{dt} = I_\nu^2 \, \cos\theta_2 \, d\Omega_2 \, d\sigma_2 \,  d\nu =
I_\nu^2 \, \cos\theta_2 \, \frac{\cos\theta_1 \, d\sigma_1}{D^2} \, d\sigma_2 \,  d\nu \, .
\ee
Since there is no absorption/emission between the infinitesimal surfaces $d\sigma_1$ and $d\sigma_2$, $dE_1 = dE_2$, and therefore $I_\nu^1 = I_\nu^2$; i.e., the specific intensity is conserved. We can repeat the calculations for the total intensity and arrive at the same conclusion.

If we integrate the specific intensity over the solid angle subtended by the source, we find the {\it spectral flux} (also called the spectral flux density or the flux density)
\be
F_\nu = \int_{\rm source} I_\nu \, \cos\theta \, d\Omega \, .
\ee
The spectral flux can be measured, for example, in W~m$^{-2}$~Hz$^{-1}$. The {\it total flux} is obtained after integrating the spectral flux over all frequencies
\be
F = \int_0^{\infty} F_\nu \, d\nu \, .
\ee
The total flux can be measured, for example, in W~m$^{-2}$.

The {\it spectral luminosity} is the total power per unit bandwidth emitted by the source at a certain frequency. It can be indicated with $L_\nu$ and can be measured, for example, in W~Hz$^{-1}$. In general, it is not straightforward to measure the spectral luminosity of an astrophysical source if its emission is not isotropic. For an isotropic source at distance $D$, the spectral luminosity is
\be
L_\nu = 4 \pi D^2 \, F_\nu \, .
\ee
The {\it total luminosity} (also called the bolometric luminosity) is obtained after integrating the spectral luminosity over all frequencies
\be
L = \int_0^{\infty} L_\nu \, d\nu \, .
\ee
The total luminosity can be measured, for example, in W.

Before concluding this subsection, it can be useful to stress a few important properties of the quantities discussed here. 
\begin{enumerate}
\item The specific intensity $I_\nu$ is independent of the distance of the source, $I_\nu = {\rm constant}$, while the spectral flux $F_\nu$ is proportional to the inverse of the square of the distance, $F_\nu \propto D^{-2}$.
\item The specific intensity $I_\nu$ can be seen as the energy flowing out of the source as well as the energy flowing into the detector, and, in general, as the energy flowing along any photon trajectory.
\item If a source is unresolved (i.e., it appears as a point-like source because its angular size is smaller than the angular resolution of the instrument used for its observation), we can measure the spectral flux $F_\nu$ but we cannot measure the specific intensity $I_\nu$.
\end{enumerate}

%%%%%%%%%%%%%%%

\subsection{Relativistic effects}\label{s-rel}

If the astrophysical source is a compact object, like a black hole or a neutron star, relativistic effects may not be ignored. Some results discussed in the previous subsection may require modifications.

If the wavelength of the radiation $\lambda$ is much smaller than the size of the source, the size of the detector, and the distance between the source and the detector, we can still use the ray-optics approximation, but in a curved spacetime the radiated energy flows along null geodesics, which may not be straight lines.

The specific intensity $I_\nu$ and the total intensity $I$ are not conserved along photons trajectories. The {\it Liouville's theorem} shows that the conserved quantity is instead $I_\nu / \nu^3$ (for the details, see Ref.~\cite{Lindquist:1966igj}). If $I_{\rm e} (\nu_{\rm e})$ is the specific intensity at the emission point at the frequency $\nu_{\rm e}$ as measured in the rest-frame of the emitter and $I_{\rm o} (\nu_{\rm o})$ is the specific intensity at the detection point at the frequency $\nu_{\rm o}$ as measured by the observer, the relation between these two specific intensities is
\be
I_{\rm o} (\nu_{\rm o}) = g^3 I_{\rm e} (\nu_{\rm e}) \, , 
\ee
where $g = \nu_{\rm o}/\nu_{\rm e}$ is the redshift factor between the detector and the emitter. If relativistic effects are negligible, $g = 1$, and we recover the result of the previous subsection.

%%%%%%%%%%%%%%%%%%%%%%%%%%%%%%%

\section{Thin accretion disk}\label{s-disk}

In this section, we present a simple model to describe thin, Keplerian accretion disks. The key-assumptions are that the disk is non-self-gravitating\footnote{This is normally a very acceptable approximation. Let us consider, for example, a 10~$M_\odot$ black hole in an X-ray binary. Its Eddington luminosity is $L_{\rm Edd} \sim 10^{39}$~erg~s$^{-1}$. Its Eddington mass accretion rate can be found from the relation $L_{\rm Edd} = \eta \dot{M}_{\rm Edd}$, where $\eta \sim 0.1$ is the radiative efficiency, and we have $\dot{M}_{\rm Edd} \sim 10^{-7}$~$M_\odot$/yr. In a typical outburst of a black hole binary, the luminosity of the source is around 10\% of its Eddington limit and the outburst lasts for about a month, so the total mass in the accretion disk is roughly $10^{-9}$~$M_\odot$ and the ratio between the mass of the disk and the mass of the black hole is of order $10^{-10}$. Even if the mass of the disk were confined in a relatively small space region (which is not the case), it could produce only a very small perturbation on the background metric. We can thus conclude that the gravitational field of the disk can be ignored. See, for instance, Ref.~\cite{Bambi:2014koa} for more details.} and that the motion of the material in the disk is determined by the gravitational field of the black hole, while other effects (for instance, the pressure of the material itself or the presence of magnetic fields) can be neglected. This implies that the motion of the accreting material is governed by the geodesic equations in the black hole spacetime.

%%%%%%%%%%%%%%%

\subsection{Motion in a stationary and axisymmetric spacetime}

We assume that the motion of the ``material''\footnote{We note that we are not considering the motion of the single particles in the accretion disk (ions and electrons) but the motion of a ``parcel'' of particles.} in the accretion disk is determined by the gravitational field of the black hole and therefore it can be derived from the Lagrangian of a point-like free particle
\be\label{eq-lag}
\mathcal{L} = \frac{1}{2} \left( g_{tt} \dot{t}^2 + 2 g_{t\phi} \dot{t} \dot{\phi} 
+ g_{rr} \dot{r}^2 + g_{\theta\theta} \dot{\theta}^2 + g_{\phi\phi} \dot{\phi}^2 \right) \, ,
\ee
where $\dot{} = d/d\tau$ and $\tau$ is the proper time of the material. Since the metric coefficients are independent of the coordinates $t$ and $\phi$, we have two constants of motion: the specific energy $E$ and the axial component of the specific angular momentum $L_z$:\footnote{We use the term ``specific'' because the Lagrangian of a point-like free particle is $\mathcal{L} = \frac{1}{2} m \left( g_{tt} \dot{t}^2 + 2 g_{t\phi} \dot{t} \dot{\phi} + g_{rr} \dot{r}^2 + g_{\theta\theta} \dot{\theta}^2 + g_{\phi\phi} \dot{\phi}^2 \right)$ and we set $m = 1$ in Eq.~(\ref{eq-lag}).} 
\be\label{eq-pt}
p_t &=& \frac{\partial \mathcal{L}}{\partial \dot{t}} = g_{tt} \dot{t} + g_{t\phi} \dot{\phi} = - E \, , \\
p_\phi &=& \frac{\partial \mathcal{L}}{\partial \dot{\phi}} = g_{t\phi} \dot{t} + g_{\phi\phi} \dot{\phi} = L_z \, . \label{eq-pphi}
\ee
From the two equations above, we can write $\dot{t}$ and $\dot{\phi}$ in terms of $E$, $L_z$, and the metric coefficients:
\be\label{eq-tdot}
\dot{t} &=& \frac{g_{\phi\phi} E + g_{t\phi} L_z}{g_{t\phi}^2 - g_{tt} g_{\phi\phi}} \, , \\
\dot{\phi} &=& - \frac{g_{t\phi} E + g_{tt} L_z}{g_{t\phi}^2 - g_{tt} g_{\phi\phi}} \, .
\label{eq-phidot}
\ee

%%%%%%%%%%%%%%%

\subsection{Infinitesimally thin disk}

Eq.~(\ref{eq-tdot}) and Eq.~(\ref{eq-phidot}) directly follow from the conservation of $E$ and $L_z$ and there are no assumptions about the orbits in the spacetime. Now we want to describe the motion of the material is in an accretion disk. This accretion disk is infinitesimally thin and perpendicular to the black hole spin axis, so the motion of the material has $\theta = \pi/2$ and $\dot{\theta} = 0$.

%%%%%%%%%%%%%%%

\subsection{Motion in the disk region ($r > r_{\rm ISCO}$)}\label{ss-disk}

The motion of the material in the disk is approximated with {\it quasi}-geodesic, equatorial, circular orbits\footnote{Here we ignore the ``quasi'' and we assume that the material in the disk follows geodesic, equatorial, circular orbits. However, the ``quasi'' is important to have accretion onto the black hole. It is indeed necessary a mechanism to transport energy and angular momentum outward, so the material of the disk can slowly inspiral onto the black hole. In reality, the {\it magnetorotational instability} is the mechanism responsible to transport energy and angular momentum outward and to permit the accretion process~\cite{Balbus:1991ay}.}, so $\dot{r} = 0$. Let us write the geodesic equations in the following form
\be\label{eq-geo}
\frac{d}{d\tau} \left( g_{\mu\nu} \dot{x}^\nu \right) = \frac{1}{2} \left( \partial_\mu g_{\nu\rho} \right) \dot{x}^\nu \dot{x}^\rho \, .
\ee
For equatorial circular orbits, $\dot{r} = \ddot{r} = \dot{\theta} = 0$, and, for $\mu = r$, Eq.~(\ref{eq-geo}) reduces to
\be
\left( \partial_r g_{tt} \right) \dot{t}^2 + 2 \left( \partial_r g_{t\phi} \right) 
\dot{t} \dot{\phi} + \left( \partial_r g_{\phi\phi} \right) \dot{\phi}^2 = 0 \, . 
\ee
The angular velocity of the material as measured by the coordinate system $(t,r,\theta,\phi)$ is $\Omega_{\rm K} = d\phi/dt = \dot{\phi}/\dot{t}$ and the previous equation becomes\footnote{The subindex ${\rm K}$ is used to indicate that this is the {\it Keplerian} angular velocity of the material in the accretion disk and to avoid confusion in the formulas between such a quantity and the infinitesimal solid angle $d\Omega$.}
\be\label{eq-omega}
\left( \partial_r g_{\phi\phi} \right) \Omega_{\rm K}^2 + 2 \left( \partial_r g_{t\phi} \right) \Omega_{\rm K} + \left( \partial_r g_{tt} \right) = 0 
\quad \Rightarrow \quad
\Omega_{\rm K} = \frac{ - \partial_r g_{t\phi} \pm \sqrt{ \left( \partial_r g_{t\phi} \right)^2 
- \left( \partial_r g_{tt} \right) \left( \partial_r g_{\phi\phi} \right) }}{\partial_r g_{\phi\phi}} \, ,
\ee
where the sign $+$ is for co-rotating orbits (orbits with angular momentum parallel to the black hole spin) and the sign $-$ is for counter-rotating orbits (orbits with angular momentum anti-parallel to the black hole spin).

Since the material of the disk follows time-like geodesics, which we are parametrizing with the proper time $\tau$, we have $g_{\mu\nu} \dot{x}^\mu \dot{x}^\nu = -1$. Since $\dot{r} = \dot{\theta} = 0$ and $\dot{\phi} = \Omega_{\rm K} \dot{t}$, we have 
\be\label{eq-tdot2}
g_{tt} \dot{t}^2 + 2 g_{t\phi} \dot{t} \dot{\phi} + g_{\phi\phi} \dot{\phi}^2 = - 1 
\quad \Rightarrow \quad
\dot{t} = \frac{1}{\sqrt{ - g_{tt} - 2 g_{t\phi} \Omega_{\rm K} - g_{\phi\phi} \Omega_{\rm K}^2}} \, .
\ee
At this point we can write the specific energy $E$ and the axial component of the specific angular momentum $L_z$ for geodesic, equatorial, circular orbits in terms of known quantities
\be\label{eq-E-c-orbit}
E &=& - \left( g_{tt} + g_{t\phi} \Omega_{\rm K} \right) \dot{t} = 
- \frac{g_{tt} + g_{t\phi} \Omega_{\rm K}}{\sqrt{ - g_{tt} - 2 g_{t\phi} \Omega_{\rm K} - g_{\phi\phi} \Omega_{\rm K}^2}} \, , \\
L_z &=& \left( g_{t\phi} + g_{\phi\phi} \Omega_{\rm K} \right) \dot{t} = 
\frac{ g_{t\phi} + g_{\phi\phi} \Omega_{\rm K} }{\sqrt{ - g_{tt} - 2 g_{t\phi} \Omega_{\rm K} - g_{\phi\phi} \Omega_{\rm K}^2}} \, .
\label{eq-L-c-orbit}
\ee

Since the motion of the material is determined by the gravitational field of the black hole (and we ignore the pressure of the fluid and possible magnetic fields), the inner edge of the accretion disk can naturally be at the radius of the {\it innermost stable circular orbit} (ISCO)~\cite{Bardeen:1972fi}. Inside the ISCO, circular orbits are unstable, so the material can quickly plunge onto the black hole. To find the ISCO radius, we need to study the stability of the geodesic, equatorial, circular orbits. From $g_{\mu\nu} \dot{x}^\mu \dot{x}^\nu = -1$ and using Eq.~(\ref{eq-tdot}) and Eq.~(\ref{eq-phidot}), we can write
\be
g_{rr} \dot{r}^2 + g_{\theta\theta} \dot{\theta}^2 = V_{\rm eff} ( r , \theta ) \, ,
\ee 
where 
\be
V_{\rm eff} = \frac{g_{\phi\phi} E^2 + 2 g_{t\phi} E L_z + g_{tt} L_z^2}{g_{t\phi}^2 - g_{tt} g_{\phi\phi}} - 1 \, .
\ee
Orbits are stable (unstable) under small perturbations along the radial and vertical directions if, respectively,
\be
\frac{\partial^2 V_{\rm eff}}{\partial r^2} < 0 \quad (> 0) \, , \qquad
\frac{\partial^2 V_{\rm eff}}{\partial \theta^2} < 0 \quad (> 0)
\ee
In the Kerr spacetime and in many other black hole spacetimes, equatorial circular orbits are always stable along the vertical direction, the ISCO radius is determined by the stability along the radial direction, and there is one ISCO radius separating stable orbits ($r > r_{\rm ISCO}$) and unstable orbits ($r < r_{\rm ISCO}$). In general, this is not guaranteed: orbits may be even vertically unstable and/or there may be more than one stable region and/or more than one unstable region~\cite{Bambi:2011vc,Bambi:2013eb}.

%%%%%%%%%%%%%%%

\subsection{Motion in the plunging region ($r_{+} < r < r_{\rm ISCO}$)}\label{ss-plunging}

At $r = r_{\rm ISCO}$, equatorial circular geodesics become unstable. We can approximate the motion of the material in the plunging region ($r_{+} < r < r_{\rm ISCO}$, where $r_{+}$ is the radial coordinate of the black hole event horizon in the equatorial plane) with equatorial geodesic orbits with specific energy $E^{\rm ISCO}$ and axial component of the specific angular momentum $L_z^{\rm ISCO}$, where $E^{\rm ISCO}$ and $L_z^{\rm ISCO}$ are $E$ and $L_z$ at $r = r_{\rm ISCO}$.

The 4-velocity of the material in the plunging region is $u^\mu = \left( \dot{t} , \dot{r} , 0 , \dot{\phi} \right)$. $\dot{t}$ and $\dot{\phi}$ are given by Eq.~(\ref{eq-tdot}) and Eq.~(\ref{eq-phidot}) for $E = E^{\rm ISCO}$, $L_z = L_z^{\rm ISCO}$, and $\theta = \pi/2$
\be\label{eq-tdot-plunging}
\dot{t} &=& \frac{g_{\phi\phi} E^{\rm ISCO} + g_{t\phi} L_z^{\rm ISCO}}{g_{t\phi}^2 - g_{tt} g_{\phi\phi}} \, , \\
\dot{\phi} &=& - \frac{g_{t\phi} E^{\rm ISCO} + g_{tt} L_z^{\rm ISCO}}{g_{t\phi}^2 - g_{tt} g_{\phi\phi}} \, .
\label{eq-phidot-plunging}
\ee
$\dot{r}$ can be inferred from $g_{\mu\nu} \dot{x}^\mu \dot{x}^\nu = -1$
\be\label{eq-rdot-plunging}
\dot{r} = - \sqrt{ \frac{-1 -g_{tt}\dot{t}^2 - 2g_{t\phi}\dot{t}\dot{\phi} - g_{\phi\phi}\dot{\phi}^2}{g_{rr}}}
= - \sqrt{ \frac{1}{g_{rr}} \left[ \frac{g_{\phi\phi} (E^{\rm ISCO})^2 + 2 g_{t\phi} 
E^{\rm ISCO} L^{\rm ISCO}_z + g_{tt} (L_z^{\rm ISCO})^2}{g_{t\phi}^2 - g_{tt} g_{\phi\phi}} - 1 \right] } \, ,
\ee
where the sign $-$ in front of the square root is chosen because the material is falling onto the black hole. It can be useful to stress that $\dot{t}$, $\dot{\phi}$, and $\dot{r}$ are functions only of the radial coordinate $r$, so we have the 4-velocity of the accreting material in the plunging region at every radial coordinate.

The presence of magnetic fields can make the model significantly more complicated. General Relativistic Magnetohydrodynamic (GRMHD) simulations can provide the most reliable description of the accretion process onto compact objects, even if current simulations still rely on a number of simplifications, so caution is necessary when we want to derive conclusions for the accretion process on astrophysical black holes, and the morphology of magnetic fields around black holes is still poorly understood. Current GRMHD studies suggest that thin accretion disks are close to be Keplerian and the simple model described in Subsection~\ref{ss-disk} can approximate well the motion of the material in the disk region~\cite{Shashank:2022xyh}. Magnetic fields can instead affect the motion of the material in the plunging region and change the location of the inner edge of the disk, which may not be $r_{\rm ISCO}$ any longer. In such a case, Eqs.~(\ref{eq-tdot-plunging}), (\ref{eq-phidot-plunging}), and (\ref{eq-rdot-plunging}) may not describe well the motion in the plunging region. The simplest models to describe the motion of the material in the plunging region in the presence of magnetic fields still assume that the motion is on the equatorial plane ($\theta = \pi/2$ and $\dot{\theta} = 0$) and consider simple modifications in the Lagrangian in Eq.~(\ref{eq-lag}) to slow down the fall of the material from the ISCO to the black hole.

%%%%%%%%%%%%%%%%%%%%%%%%%%%%%%%

\section{Corona}\label{s-corona}

This section is devoted to study how the corona illuminates the accretion disk.

%%%%%%%%%%%%%%%

\subsection{Coronal spectrum}

We start assuming that the corona is a point-like source. An extended corona can be easily described as the combination of a number of point-like coronae.

The {\it photon spectrum} of the corona in its rest-frame can be approximated by a power law with a high-energy cutoff $E_{\rm max}$ and a low-energy cutoff $E_{\rm min}$, so we can write
\be\label{eq-corona-spectrum}
\frac{dN_{\rm c}}{dt_{\rm c} dE_{\rm c}} =
\left\{
\begin{array}{cl}
K E^{-\Gamma}_{\rm c} & \quad E_{\rm min} < E_{\rm c} < E_{\rm max} \\
0 & \quad \text{otherwise}
\end{array}
\right. \, ,
\ee
where $N_{\rm c}$ is the photon number, $K$ is a normalization constant, $\Gamma$ is the photon index, and the subindex ${\rm c}$ indicates that a quantity is evaluated in the rest-frame of the corona. The spectral luminosity of the corona is
\be
\frac{dL_{\rm c}}{dE_{\rm c}} = E_{\rm c} \frac{dN_{\rm c}}{dt_{\rm c} dE_{\rm c}} \, , 
\ee
and the total luminosity of the corona is
\be
L_{\rm c} = \int E_{\rm c} \frac{dN_{\rm c}}{dt_{\rm c} dE_{\rm c}} \, dE_{\rm c}
= \int_{E_{\rm min}}^{E_{\rm max}} K E_{\rm c}^{-\Gamma+1} \, dE_{\rm c} 
= \left\{
\begin{array}{cl}
\frac{K}{2-\Gamma} \left[ E_{\rm max}^{2-\Gamma} - E_{\rm min}^{2-\Gamma} \right] & \quad \Gamma \neq 2 \\
\\
K \ln \left( \frac{E_{\rm max}}{E_{\rm min}} \right) & \quad \Gamma = 2
\end{array}
\right. \, .
\ee
If we know the total luminosity of the corona and its spectrum, we can fix the normalization constant $K$.

The spectrum of the corona detected far from the source is not exactly that in Eq.~(\ref{eq-corona-spectrum}) because it is redshifted. The redshift factor is
\be
g = \frac{E_{\rm o}}{E_{\rm c}} = \frac{ - u_{\rm o}^\mu k_\mu }{ - u_{\rm c}^\mu k_\mu } \, ,
\ee
where $u_{\rm o}^\mu = ( 1 , 0 , 0 , 0 )$ is the 4-velocity of the distant observer, $u_{\rm c}^\mu$ is the 4-velocity of the point-like corona, and $k_\mu = (k_t , k_r , k_\theta , k_\phi)$ is the conjugate 4-momentum of the photons emitted by the corona and detected by the distant observer. $k_\mu$ should be evaluated at the detection point in the numerator and at the corona in the denominator. The photon spectrum of the corona detected by the distant observer is
\be\label{eq-corona-spectrum-observer}
\frac{dN_{\rm o}}{dt_{\rm o} dE_{\rm o}} =
\left\{
\begin{array}{cl}
K' E^{-\Gamma}_{\rm o} & \quad {E'}_{\rm min} < E_{\rm o} < {E'}_{\rm max} \\
0 & \quad \text{otherwise}
\end{array}
\right. \, ,
\ee
where ${E'}_{\rm max} = g E_{\rm max}$ and ${E'}_{\rm min} = g E_{\rm min}$ are, respectively, the redshifted high-energy cutoff and the redshifted low-energy cutoff, $K' = g^\Gamma K$, and the subindex ${\rm o}$ indicates that a quantity is evaluated in the rest-frame of the observer. The photon number is an invariant. If the corona emits photons from time $t_{\rm c}$ to time $t_{\rm c} + \Delta t_{\rm c}$ in the corona rest-frame, the total number of photons is (for $\Gamma \neq 1$)
\be
N_{\rm c} = \int_{t_{\rm c}}^{t_{\rm c} + \Delta t_{\rm c}} dt_{\rm c} 
\int_{E_{\rm min}}^{E_{\rm max}} dE_{\rm c} \, K E_{\rm c}^{-\Gamma}
=  \frac{\Delta t_{\rm c} K}{1-\Gamma} \left[ E_{\rm max}^{1-\Gamma} - E_{\rm min}^{1-\Gamma} \right] \, .
\ee 
The total number of photons detected by the distant observer is
\be
N_{\rm o} &=& \int_{t_{\rm o}}^{t_{\rm o} + \Delta t_{\rm o}} dt_{\rm o} 
\int_{{E'}_{\rm min}}^{{E'}_{\rm max}} dE_{\rm o} \, K' E_{\rm o}^{-\Gamma}
=  \frac{\Delta t_{\rm o} K'}{1-\Gamma} \left[ {E'}_{\rm max}^{1-\Gamma} - {E'}_{\rm min}^{1-\Gamma} \right] \nonumber\\
&=& \frac{\left( g^{-1} \Delta t_{\rm c} \right) \left( g^\Gamma K \right)}{1-\Gamma} 
\left[ \left( g^{1-\Gamma} \right) E_{\rm max}^{1-\Gamma} - \left( g^{1-\Gamma} \right) E_{\rm min}^{1-\Gamma} \right]
= N_{\rm c} \, .
\ee

%%%%%%%%%%%%%%%

\subsection{Illumination of the accretion disk}

To calculate how to corona illuminates the disk, we fire $\mathcal{N}$ photons from the corona to the disk, where $\mathcal{N}$ is a sufficiently large number to meet our requirements of accuracy for the final result. If the corona emission is isotropic in its rest-frame, we consider the rest-frame of the corona and we fire isotropically these $\mathcal{N}$ photons. If this is not the case, we consider the rest-frame of the corona and we fire $\mathcal{N}$ photons according to the specific angular emission law of the corona such that any photon trajectory can represent the same number of photons per unit time and unit energy. The photon spectrum for every emission direction (ray) is 
\be
\frac{dN_{\rm c}}{dt_{\rm c} dE_{\rm c}} =
\left\{
\begin{array}{cl}
\tilde{K} E^{-\Gamma}_{\rm c} & \quad E_{\rm min} < E < E_{\rm max} \\
0 & \quad \text{otherwise}
\end{array}
\right. \, ,
\ee
where $\tilde{K} = K/\mathcal{N}$.

\begin{figure}[b]
\centering
\includegraphics[width=0.45\linewidth,trim={0cm 0cm 0cm 4.0cm},clip]{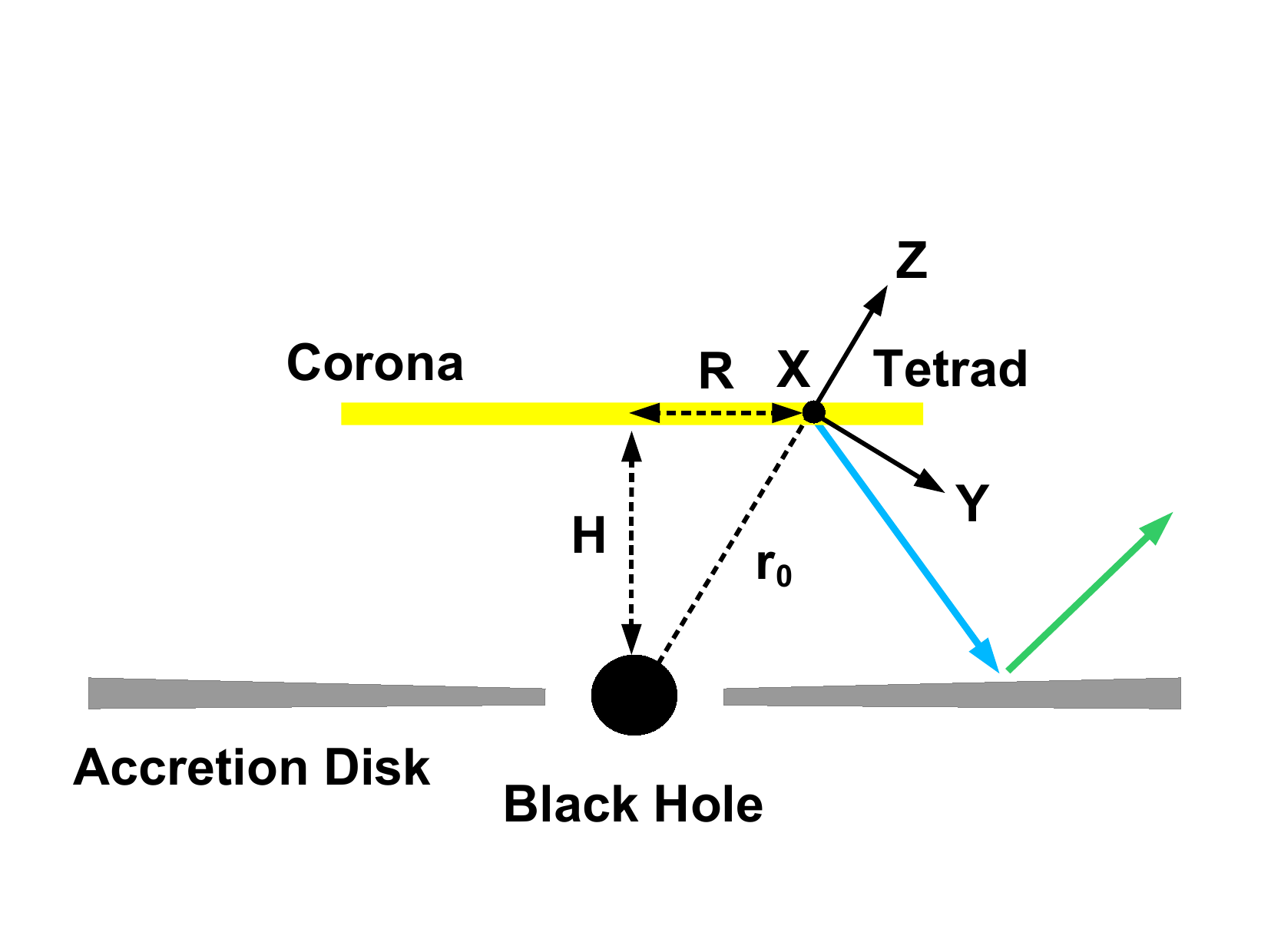}
\vspace{-1.0cm}
\caption{Rotating disk-like corona: an extended corona can be thought of as the combination of a number of point-like coronae. For every point-like corona, we consider its locally Minkowskian reference frame and we fire $\mathcal{N}$ photons. We calculate the null geodesics from the point-like corona to the accretion disk in order to calculate how the point-like corona illuminates the disk. See the text for more details.}
\label{f-diskcorona}
\end{figure}

As an example, let us assume that the corona is an infinitesimally thin disk above the accretion disk and the black hole, as shown in Fig.~\ref{f-diskcorona}~\cite{Riaz:2020svt}. The system is perfectly axisymmetric, so the central axis of the corona coincides with that of the accretion disk and with the rotational axis of the black hole. The corona may rotate with angular frequency $\tilde{\omega}$ as measured in the spherical-like coordinates $(t,r,\theta,\phi)$. Since the system is axisymmetric, we can restrict our calculations to a certain radial direction. We consider a number of point-like coronae along such a radial direction and, as shown in Fig.~\ref{f-diskcorona}, we indicate with $R$ the radial coordinate of every point-like corona from the central axis. The spacing of the point-like coronae can be used to regulate the intensity profile of our disk-like corona. For example, if we assume that the surface of the corona has constant luminosity, we need a point-like corona for every equal-area annulus. In the Newtonian limit, the area of the annulus of radius $R$ and width $\Delta R$ is $2 \pi \, R \, \Delta R$, and therefore we would need to distribute the point-like coronae with a separation $\Delta R \propto 1/R$. If we employ $\Delta R = {\rm constant}$, it is equivalent to assume that the intensity profile of the corona scales as $1/R$.

To write the photon initial conditions, it is convenient to choose the locally Minkowskian reference frames of the point-like coronae (see Appendix~\ref{a1}). If $u_j^\mu$ is the 4-velocity of the point-like corona $j$, $u_j^r = u_j^\theta = 0$ (because the corona rotates about the black hole spin axis) and $u_j^\phi = \tilde{\omega} u_j^t$ by definition. From $g_{\mu\nu} u_j^\mu u_j^\nu = -1$, we find $u_j^t$ 
\be
u_j^t = \frac{1}{\sqrt{ -g_{tt} - 2 \tilde{\omega} g_{t\phi} - \tilde{\omega}^2 g_{\phi\phi}}} \, .
\ee
The time-like tetrad basis vector $E^\mu_{(T)}$ is $u_j^\mu$. We choose the orientation of the space-like tetrad basis vectors as shown in Fig.~\ref{f-diskcorona} with $\hat{r} = \hat{Z}$ and $\hat{\theta} = \hat{Y}$. The expression for $E^\mu_{(X)}$ can be inferred from the conditions $g_{\mu\nu} E^\mu_{(X)} E^\nu_{(X)} = 1$ and $g_{\mu\nu} E^\mu_{(X)} E^\nu_{(T)} = g_{\mu\nu} E^\mu_{(X)} E^\nu_{(Y)} = g_{\mu\nu} E^\mu_{(X)} E^\nu_{(Z)} = 0$. In the end, the tetrad of the orthogonal basis vectors associated to the locally Minkowskian reference frame of the point-like corona $j$ has the following form
\be\label{eq-tetrad-corona}
E^\mu_{(T)} = u_j^t 
\left(
\begin{array}{c}
1 \\
0 \\
0 \\ 
\tilde{\omega} 
\end{array}
\right) , \;\;
%%%
E^\mu_{(X)} = 
\frac{u_j^t}{\sqrt{g^2_{t\phi} - g_{tt} g_{\phi\phi} }}
\left(
\begin{array}{c}
g_{t\phi} + \tilde{\omega} g_{\phi\phi} \\
0 \\
0 \\ 
-g_{tt} - \tilde{\omega} g_{t\phi}
\end{array}
\right) , \;\; 
%%%
E^\mu_{(Y)} = \frac{1}{\sqrt{g_{\theta\theta}}}
\left(
\begin{array}{c}
0 \\
0 \\
1 \\ 
0 
\end{array}
\right) , \;\;
%%%
E^\mu_{(Z)} = \frac{1}{\sqrt{g_{rr}}}
\left(
\begin{array}{c}
0 \\
1 \\
0 \\ 
0 
\end{array}
\right) .
\ee

For every point-like corona, we fire $\mathcal{N}$ photons and check if and where these photons hit the accretion disk. In the locally Minkowskian reference frame of the point-like corona, the initial 4-momentum of the photons can be written as
\be
k^{(\alpha)}_0 = 
\left(
\begin{array}{c}
E \\
E \sin\chi \cos\psi \\
E \sin\chi \sin\psi \\
E \cos\chi
\end{array}
\right) \, ,
\ee
where $E$ is the photon energy and $\chi \in [0 ; \pi]$ and $\psi \in [0 ; 2\pi]$ are the polar angles in the locally Minkowskian reference frame of the point-like corona. If the emission of the point-like corona is isotropic, we can consider a grid of constant $\Delta\left(\cos\chi\right)$ and $\Delta\psi$ and fire a photon from every point of the grid\footnote{If the emission is isotropic, we can divide the sky of every emission point in small solid angles $\Delta\Omega = {\rm constant}$ and fire a photon from each of these directions. Since $d\Omega = \sin\chi \, d\chi \, d\psi$, we need a grid of constant $\Delta\left(\cos\chi\right)$ and $\Delta\psi$.}. In the spherical-like coordinates $(t,r,\theta,\phi)$, the initial conditions for the photon position will be like
\be\label{eq-in-pos}
t_0 = 0 \, , \quad r_0 = \sqrt{H^2 + R^2} \, , \quad \theta_0 = \arctan \left( \frac{R}{H} \right) \, , \quad \phi_0 = 0 \, .
\ee
The initial conditions for the photon 4-momentum will be like 
\be
k^\mu_0 = k^{(\alpha)}_0 E^\mu_{(\alpha)} \, .
\ee
In our case, with the tetrad in~(\ref{eq-tetrad-corona}), we have
\be\label{eq-in-mom}
k^t_0 = k^{(T)}_0 E^t_{(T)} + k^{(X)}_0 E^t_{(X)} \, , \quad
k^r_0 = k^{(Z)}_0 E^r_{(Z)} \, , \quad
k^\theta_0 = k^{(Y)}_0 E^\theta_{(Y)} \, , \quad
k^\phi_0 = k^{(T)}_0 E^\phi_{(T)} + k^{(X)}_0 E^\phi_{(X)} \, .
\ee
With the initial conditions in (\ref{eq-in-pos}) and (\ref{eq-in-mom}), we solve the geodesic equations in the coordinate system $(t,r,\theta,\phi)$. We calculate the trajectory of every photon of the grid of the point-like corona and then we repeat these calculations for every point-like corona.

%%%%%%%%%%%%%%%

\subsection{Spectral flux illuminating the disk}

We divide the accretion disk into annuli (radial bins). The annulus $i$ has radial coordinate $r_i$ and width $\Delta r_i$. We fire photons from the point-like corona $j$ to the disk, as discussed in the previous subsection. If the ray $k$ ($k = 1, ... , \mathcal{N}$) hits the annulus $i$, it deposits the following energy on the annulus $i$ with photons of energy $E_{\rm d}$
\be
d\varepsilon_{ijk} = E_{\rm d} dN \, ,
\ee 
where $dN$ is the photon number, which is conserved along the photon path because we are in vacuum and there is no photon emission or absorption. The photon number of a ray in the rest-frame of the corona is
\be\label{eq-dNijk}
dN = \tilde{K} E^{-\Gamma}_{\rm c} \, d t_{\rm c} \, d E_{\rm c} \, ,
\ee  
The redshift factor between the emission point (the point-like corona $j$) and the incident point (the annulus $i$) of the ray $k$ is
\be
g_{ijk} = \frac{E_{\rm d}}{E_{\rm c}} = \frac{- u^\mu_i k_\mu}{- u^\mu_j k_\mu} 
= \left[ \frac{1 - b_k \, \Omega_{\rm K}}{\sqrt{-g_{tt} - 2 \Omega_{\rm K} g_{t\phi} - \Omega_{\rm K}^2 g_{\phi\phi}}} \right]_i
\left[ \frac{\sqrt{-g_{tt} - 2 \tilde{\omega} g_{t\phi} - \tilde{\omega}^2 g_{\phi\phi}}}{1 - b_k \, \tilde{\omega}} \right]_j \, ,
\ee
where $b_k = - k_\phi / k_t$ is a constant of motion of the ray $k$, the subindices $i$ and $j$ in the last expression are used to indicate that the first term is evaluated at the annulus $i$ and the second term is evaluated at the point-like corona $j$, $\Omega_{\rm K}$ is the angular velocity of the material in the disk given in Eq.~(\ref{eq-omega}), and $\tilde{\omega}$ is the angular velocity of the point-like corona. We can thus rewrite Eq.~(\ref{eq-dNijk}) as
\be
dN = \tilde{K} \, g^\Gamma_{ijk} \, E^{-\Gamma}_{\rm d} \, d t_{\rm d} \, d E_{\rm d} \, .
\ee
The spectral flux on the annulus $i$ produced by the ray $k$ of the point-like corona $j$ is
\be
F_{X , ijk} = \frac{d\varepsilon_{ijk}}{d\sigma_i \, d t_{\rm d} \, d E_{\rm d}} 
= \frac{\tilde{K} \, g^\Gamma_{ijk} \, E^{-\Gamma+1}_{\rm d}}{A_i} \, ,
\ee 
where $A_i$ is the {\it proper area} of the annulus $i$ (see Appendix~\ref{a2})
\be\label{eq-properarea}
A (r_i , \Delta r_i)
= 2\pi \Delta r_i
\left[ \sqrt{\frac{g_{rr} \left( g^2_{t\phi} - g_{tt} g_{\phi\phi} \right)}{-g_{tt} - 2 g_{t\phi} \Omega_{\rm K} - g_{\phi\phi} \Omega_{\rm K}^2}} \right]_{r = r_i , \theta=\pi/2}
 \, .
\ee

If the ray $k$ hits the plunging region and we want to include the emission of the plunging region in the calculations, we proceed in the same way but we need to use the 4-velocity of the material in the plunging region for $u^\mu_i$ (see Subsection~\ref{ss-plunging}) and the proper area of an annulus in the plunging region for $A (r_i , \Delta r_i)$\footnote{For the proper area of an annulus in the plunging region, we can consider a locally non-rotating observer as described in Appendix~\ref{a2} and evaluate the Lorentz factor $\gamma$ with Eq.~(\ref{eq-a2-gamma}) by using the 4-velocity of the material in the plunging region for $u^\mu$.}.

The spectral flux on the annulus $i$ produced by the disk-like corona is obtained by summing over all rays for every point-like corona and then over all point-like coronae
\be\label{eq-sp-flux-diskcorona}
F_{X , i} = \sum_j \sum_k F_{X , ijk} \, .
\ee
The total flux illuminating the annulus $i$ by the disk-like corona is
\be\label{eq-tot-flux-diskcorona}
F_i = \int F_{X , i} \, dE_{\rm d} \, .
\ee
Note that the ionization parameter of the annulus $i$ is
\be
\xi_i = \frac{4 \pi F_i}{n_{{\rm e},i}} \, ,
\ee
where $n_{{\rm e},i}$ is the electron density of the annulus $i$.

We repeat these calculation for every annulus $i$ of the accretion disk and we find the spectral flux and the total flux at every radial coordinate: $F_X = F_X (r)$ and $F = F (r)$.

If we are interested only in the X-ray flux illuminating the disk (0.1-100~keV) and $g_{ijk} E_{\rm min} < 0.1$~keV and $g_{ijk} E_{\rm max} > 100$~keV, Eqs.~(\ref{eq-sp-flux-diskcorona}) and (\ref{eq-tot-flux-diskcorona}) become, respectively, (for $\Gamma \neq 2$)
\be
F_{X , i} (E_{\rm d}) &=& \frac{\tilde{K} \, E^{-\Gamma+1}_{\rm d}}{A_i} \sum_j \sum_k g^\Gamma_{ijk} \, , \\
F_i &=& \frac{\tilde{K}}{A_i \left( 2 - \Gamma \right)} \left[ E^{2 - \Gamma}_{\rm high} - E^{2 - \Gamma}_{\rm low} \right] \sum_j \sum_k g^\Gamma_{ijk} \, ,
\ee
where $E_{\rm low} = 0.1$~keV and $E_{\rm high} = 100$~keV.

If the photon index $\Gamma$ changes over the corona, Eq.~(\ref{eq-dNijk}) becomes
\be
dN = \tilde{K} E^{-\Gamma_j}_{\rm c} \, d t_{\rm c} \, d E_{\rm c} \, ,
\ee
where $\Gamma_j$ is the photon index of the point-like corona $j$. We can proceed in the same way and the spectral flux and the total flux of the disk-like corona will be still given by Eqs.~(\ref{eq-sp-flux-diskcorona}) and (\ref{eq-tot-flux-diskcorona}).

%%%%%%%%%%%%%%%%%%%%%%%%%%%%%%%

\section{Non-relativistic reflection spectrum}

The {\it non-relativistic} reflection spectrum is the reflection spectrum in the rest-frame of the material of the disk. In the previous section, we have calculated the spectrum of the radiation illuminating the disk. We can plug this spectrum in a reflection model and get as output the non-relativistic reflection spectrum. In General Relativity, the atomic physics near a black hole is the same as the atomic physics in our laboratories on Earth. This is because in General Relativity the gravitational interaction is only described by the metric tensor $g_{\mu\nu}$ and locally the laws of non-gravitational physics are those of Special Relativity: indeed we can always perform a coordinate transformation and find a locally inertial reference frame (see, for example, Subsection~6.4.2 in Ref.~\cite{Bambi:2018drb}). In theories beyond General Relativity, this may not be true\footnote{In models in which gravity does not universally couple to matter, we can have, for example, the phenomenon of variation of ``fundamental'' constants, namely constants like the fine structure constant $\alpha$, the electron mass $m_{\rm e}$, etc. may not be actual constants and may change in space and/or time~\cite{Bambi:2022lhq}. For example, their value in the strong gravitational field of a black hole may be different from their value in our laboratories on Earth. In such a context, the atomic physics may depend on the gravitational field.}. If we know the spectrum of the radiation at every radius of the accretion disk, we can calculate the non-relativistic reflection spectrum at every radial coordinate. Since in General Relativity and in any other metric theory of gravity these calculations involve only atomic physics, they will not be discussed here.

%%%%%%%%%%%%%%%%%%%%%%%%%%%%%%%

\section{Returning radiation}

\begin{figure}[b]
\centering
\includegraphics[width=0.45\linewidth]{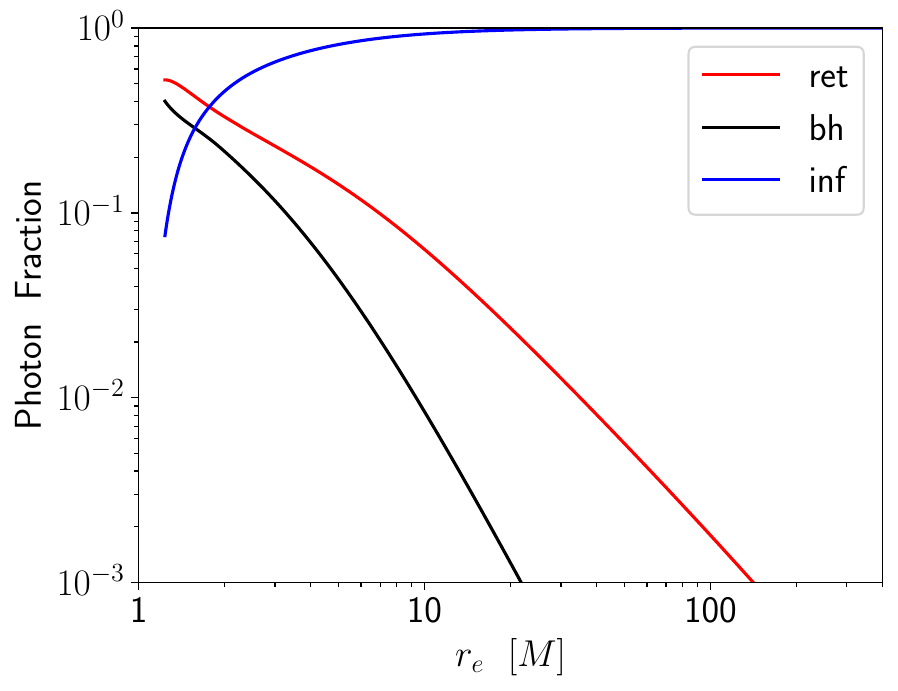}
\vspace{-0.3cm}
\caption{Fraction of photons that at every emission radius $r_{\rm e}$ return to the accretion disk (returning radiation, red curve), fall onto the black hole or the plunging region (black curve), or escape to infinity (blue curves). We assume that the compact object is a Kerr black hole with spin parameter $a_* = 0.998$ and that the accretion disk is Keplerian, infinitesimally thin, and perpendicular to the black hole spin axis. We ignore the radiation from the plunging region and the three curves start from $r_{\rm e} = r_{\rm ISCO}$ on the left.}
\label{f-rr}
\end{figure}

The {\it returning radiation} (or self-irradiation) is the radiation emitted by the disk and returning to the disk because of the phenomenon of light bending. Fig.~\ref{f-rr} shows the fractions of photons that at every emission radius return to the disk (red curve), fall onto the black hole or the plunging region (black curve), or escape to infinity (blue curve). The plot refers to the case of an infinitesimally thin, Keplerian disk in the equatorial plane of a Kerr black hole with spin parameter $a_* = 0.998$. As we can see from Fig.~\ref{f-rr}, the fraction of photons returning to the accretion disk is relevant only at very small radii, say $r < 2$~$M$, while at larger radii most of the radiation can escape to infinity~\cite{Mirzaev:2024fgd}.

To include the returning radiation in our calculations, we can consider a number of points on the accretion disk along a certain radial direction and fire photons from every point to study if and where they return to the accretion disk. At every emission point, we consider the locally Minkowskian reference frame associated to the material of the accretion disk. In the disk region $r > r_{\rm ISCO}$, the tetrad of the orthogonal basis vectors and its inverse are those reported in Eq.~(\ref{eq-tetrad}) and Eq.~(\ref{eq-tetrad-inverse}) with $\omega = \Omega_{\rm K}$
\be
&&E^\mu_{(T)} = u^t 
\left(
\begin{array}{c}
1 \\
0 \\
0 \\ 
\Omega_{\rm K} 
\end{array}
\right) , \;
%%%
E^\mu_{(X)} = 
\frac{1}{\sqrt{g_{rr}}}
\left(
\begin{array}{c}
0 \\
1 \\
0 \\ 
0 
\end{array}
\right) , \;
%%%
E^\mu_{(Y)} = 
\frac{u^t}{\sqrt{g^2_{t\phi} - g_{tt} g_{\phi\phi} }}
\left(
\begin{array}{c}
g_{t\phi} + \Omega_{\rm K} g_{\phi\phi} \\
0 \\
0 \\ 
-g_{tt} - \Omega_{\rm K} g_{t\phi}
\end{array}
\right) , \;
%%%
E^\mu_{(Z)} = 
\frac{1}{\sqrt{g_{\theta\theta}}}
\left(
\begin{array}{c}
0 \\
0 \\
1 \\ 
0 
\end{array}
\right) ,
\\
&& E^{(T)}_\mu = - u^t 
\left(
\begin{array}{c}
g_{tt} + \Omega_{\rm K} g_{t\phi} \\
0 \\
0 \\ 
g_{t\phi} + \Omega_{\rm K} g_{\phi\phi}
\end{array}
\right) , \;
%%%
E^{(X)}_\mu = 
\sqrt{g_{rr}}
\left(
\begin{array}{c}
0 \\
1 \\
0 \\ 
0 
\end{array}
\right) , \;
%%%
E^{(Y)}_\mu = 
u^t \sqrt{g^2_{t\phi} - g_{tt} g_{\phi\phi} }
\left(
\begin{array}{c}
- \Omega_{\rm K} \\
0 \\
0 \\ 
1
\end{array}
\right) , \;
%%%
E^{(Z)}_\mu = 
\sqrt{g_{\theta\theta}}
\left(
\begin{array}{c}
0 \\
0 \\
1 \\ 
0 
\end{array}
\right) , \qquad
\ee
and $u^t$ is given in Eq.~(\ref{eq-tdot2}) because we are considering the material in the disk.

In the plunging region $r_{+} < r < r_{\rm ISCO}$, $E^\mu_{(T)} = ( u^t , u^r , 0 , u^\phi)$, where $u^t$, $u^r$, and $u^\phi$ are given, respectively, by Eq.~(\ref{eq-tdot-plunging}), Eq.~(\ref{eq-rdot-plunging}), and Eq.~(\ref{eq-phidot-plunging}). $E_{(Z)}^\mu = ( 0 , 0 , 1/\sqrt{g_{\theta\theta}} , 0)$ as in the disk region, because the material is still assumed to be confined on the equatorial plane. $E_{(X)}^\mu$ and $E_{(Y)}^\mu$ can be found from the condition $g_{\mu\nu} E_{(\alpha)}^\mu E_{(\beta)}^\nu = \eta_{(\alpha)(\beta)}$.

In such locally Minkowskian reference frames, the 4-momentum of the photons have the form
\be
k^{(\alpha)}_0 = 
\left(
\begin{array}{c}
k^{(T)}_0 \\
k^{(X)}_0 \\
k^{(Y)}_0 \\
k^{(Z)}_0
\end{array}
\right) =
\left(
\begin{array}{c}
E \\
E \sin\vartheta_{\rm e} \cos\varphi_{\rm e} \\
E \sin\vartheta_{\rm e} \sin\varphi_{\rm e} \\
E \cos\vartheta_{\rm e}
\end{array}
\right) \, ,
\ee
where $E$ is the photon energy and $\vartheta_{\rm e} \in [0 ; \pi/2]$ and $\varphi_{\rm e} \in [0 ; 2\pi]$ are the polar angles in the rest-frame of the material of the disk. If the emission is isotropic, we can consider a grid of constant $\Delta\left(\cos\vartheta_{\rm e}\right)$ and $\Delta\varphi_{\rm e}$ and fire a photon from every point of the grid. The photon 4-momenta in the coordinate system $(t,r,\theta,\phi)$ can be obtained from
\be
k^\mu_0 = E^\mu_{(\alpha)} k^{(\alpha)}_0 \, .
\ee
We can then solve the geodesic equations in the coordinate system $(t,r,\theta,\phi)$ to calculate the photon trajectories and study which trajectories return to the disk. For the photons that return to the disk, we can calculate the redshift factor
\be
g = \frac{E_2}{E_1} = \frac{ - u_2^\mu k_\mu }{ - u_1^\mu k_\mu }
\ee
where the subindices 1 and 2 refer, respectively, to the emission point and the absorption point. In the numerator, we have the 4-velocity of the material in the disk and the photon 4-momentum at the absorption point. In the denominator, we have the 4-velocity of the material in the disk and the photon 4-momentum at the emission point. For $r \ge r_{\rm ISCO}$, we use the 4-velocity of the material in the disk region (Subsection~\ref{ss-disk}) 
\be
u^\mu k_\mu = 
\frac{1 - b \, \Omega_{\rm K}}{\sqrt{ - g_{tt} - 2 g_{t\phi} \Omega_{\rm K} - g_{\phi\phi} \Omega_{\rm K}^2 }} \, ,
\ee
where $b = - k_\phi/k_t$. For $r < r_{\rm ISCO}$, we use the 4-velocity of the material in the plunging region (Subsection~\ref{ss-plunging})
\be
u^\mu k_\mu &=& 
k_t \frac{g_{\phi\phi} E^{\rm ISCO} + g_{t\phi} L_z^{\rm ISCO}}{g_{t\phi}^2 - g_{tt} g_{\phi\phi}}  
- k_\phi \frac{g_{t\phi} E^{\rm ISCO} + g_{tt} L_z^{\rm ISCO}}{g_{t\phi}^2 - g_{tt} g_{\phi\phi}} \nonumber\\
&& - k_r \sqrt{ \frac{1}{g_{rr}} \left[ \frac{g_{\phi\phi} (E^{\rm ISCO})^2 + 2 g_{t\phi} 
E^{\rm ISCO} L^{\rm ISCO}_z + g_{tt} (L_z^{\rm ISCO})^2}{g_{t\phi}^2 - g_{tt} g_{\phi\phi}} - 1 \right] } \, .
\ee
We can proceed as in the case of the illumination of the disk by the corona discussed in Section~\ref{s-corona}. We divide the disk into annuli and we calculate the spectral flux for every annulus as in Eq.~(\ref{eq-sp-flux-diskcorona}). Note that the total spectral flux illuminating the disk eventually will be the sum of the direct radiation from the corona, the returning radiation of the thermal component of the disk, and the returning radiation of the reflection component~\cite{Mirzaev:2024qcu}.

%%%%%%%%%%%%%%%%%%%%%%%%%%%%%%%

\section{Relativistic reflection spectrum}\label{s-relref}

Once we have the non-relativistic reflection spectrum at every radial coordinate of the accretion disk, we can proceed to calculate the {\it relativistic} reflection spectrum, namely the reflection spectrum of the whole disk as seen by an observer far from the source. The observed spectral flux is   
\be\label{eq-oflux}
F_{\rm o} ( E_{\rm o} ) = \int_{\rm source} I_{\rm o} ( E_{\rm o} , X , Y ) \, d\Omega
= \frac{1}{D^2} \int_{\rm source} g^3 \, I_{\rm e} ( E_{\rm e} , r_{\rm e} , \vartheta_{\rm e} ) \, dX dY
\ee
where the subindices ${\rm o}$ and ${\rm e}$ are used to indicate quantities measured in the rest-frame of the distant observer and in the rest-frame of the emitter, respectively. $d\Omega = dX dY / D^2$ is the infinitesimal solid angle in the sky of the distant observer, $X$ and $Y$ are the Cartesian coordinates in the plane of the distant observer\footnote{While current X-ray observatories do not have the necessary angular resolution to resolve the accretion disk of a black hole and the system appears as a point-like source, we have to consider an ideal observer with an excellent angular resolution in order to calculate the theoretical spectrum of an accretion disk within our model.}, and $D$ is the distance between the observer and the source. $g = E_{\rm o}/E_{\rm e}$ is the redshift factor and $I_{\rm o} = g^3 I_{\rm e}$ follows from Liouville's theorem (see Subsection~\ref{s-rel}). $r_{\rm e}$ is the emission radius in the accretion disk and $\vartheta_{\rm e}$ is the emission angle in the rest-frame of the material of the disk, namely the angle between the normal to the disk and the photon emission direction measured in the rest-frame of the material (which is different, in general, from the inclination angle of the disk $i$, namely the angle between the normal to the disk and the line of sight of the distant observer, because of the phenomenon of light bending). The plane of the distant observer is assumed to be perpendicular to the straight line connecting the source and the observer, so the angle $\theta$ in the definition of the specific intensity (see Subsection~\ref{ss-basic}) is $0$ and $\cos\theta = 1$.

We consider a grid in the image plane of the distant observer. From every point of the grid, we fire a photon and we calculate its trajectory backwards in time from the detection point on the image plane of the distant observer to the emission point on the accretion disk. When the photon hits the disk, say at the radial coordinate $r_{\rm e}$, we calculate the redshift factor $g$ and the emission angle $\vartheta_{\rm e}$. At this point, for every pixel with Cartesian coordinates $(X,Y)$ and area $dXdY$ of the image of the disk on the plane of the distant observer, we have $g$ and $I_{\rm e}$, so we can calculate the integral in Eq.~(\ref{eq-oflux}) and determine the observed flux density. The next subsections present the calculations of the photon initial conditions, redshift factor $g$, and emission angle $\vartheta_{\rm e}$.

%%%%%%%%%%%%%%%

\subsection{Photon initial conditions}\label{ss-initialcond}

\begin{figure}[b]
\centering
\includegraphics[width=0.6\linewidth]{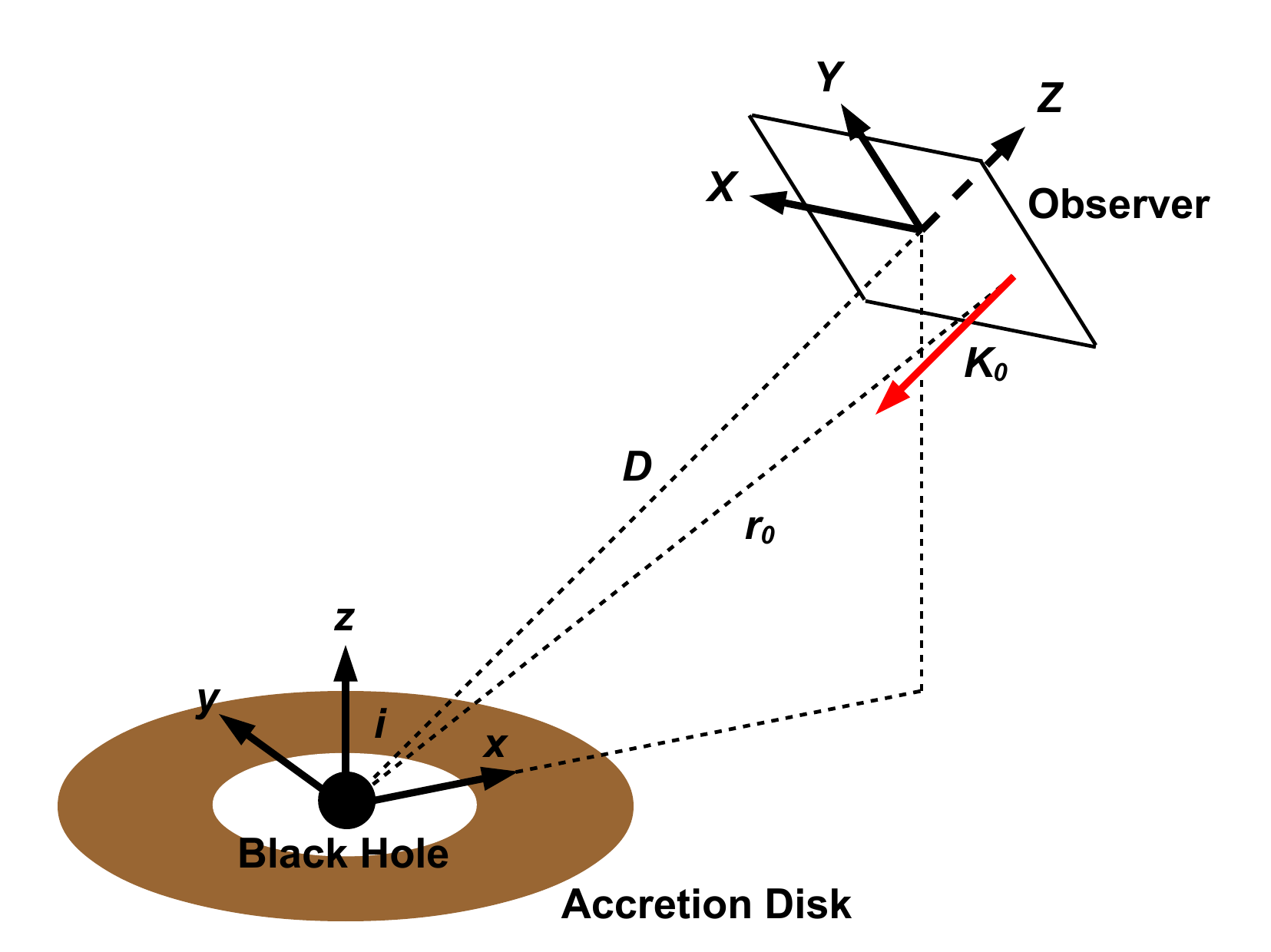}
\vspace{-0.3cm}
\caption{The Cartesian coordinates $(x,y,z)$ are centered at the black hole. The Cartesian coordinates $(X,Y,Z)$ are the coordinate system of the distant observer. $D$ is the distance between the black hole and the distant observer. $i$ is the inclination angle between the black hole spin axis and the line of sight of the distant observer. ${\bf K}_0 = (0 , 0 , -K_0)$ is the initial 3-momentum of a photon in the coordinate system of the distant observer. See the text for more details.}
\label{f-setup}
\end{figure}

To write the photon initial conditions, we consider the system sketched in Fig.~\ref{f-setup}. We have a black hole, an accretion disk on the equatorial plane of the black hole and perpendicular to the black hole spin axis, and an observer far from the black hole. $D$ is the distance between the black hole and the observer and $i$ is the inclination angle of the disk, namely the angle between the black hole spin axis and the line of sight of the distant observer. The Cartesian coordinates $(x,y,z)$ are centered at the black hole as shown in Fig.~\ref{f-setup}, so the $xy$-plane coincides with the plane of the accretion disk and the $z$-axis coincides with the black hole spin axis. The distant observer has the Cartesian coordinates $(X,Y,Z)$, where the $XY$-plane coincides with the image plane of the observer and the $Z$-axis coincides with the straight line connecting the black hole and the observer. The Cartesian coordinates $(x,y,z)$ are related to the Cartesian coordinates $(X,Y,Z)$ by the following relations
\be
x &=& D \sin i - Y \cos i + Z \sin i \, , \nonumber\\
y &=& X \, , \nonumber\\
z &=& D \cos i + Y \sin i + Z \cos i \, .
\ee

The metric of our spacetime is in spherical-like coordinates $(t,r,\theta,\phi)$. Far from the black holes, the spatial coordinates $(r,\theta,\phi)$ reduces to the usual spherical coordinates in flat spacetime and are related to $(x,y,z)$ by
\be
r &=& \sqrt{x^2 + y^2 + z^2} \, , \nonumber\\
\theta &=& \arccos \left( \frac{z}{r} \right) \, , \nonumber\\
\phi &=& \arctan \left( \frac{y}{x} \right) \, .
\ee

For the distant observer, a generic photon has initial position of the form $(X_0 , Y_0 , 0)$ and initial 3-momentum of the form ${\bf K}_0 = (0 , 0 , -K_0)$ because the photon trajectory is perpendicular to the $XY$-plane. In the coordinate system $(t,r,\theta,\phi)$ the initial position of the photon is
\be\label{eq-trtp-initial}
t_0 &=& 0 \, , \nonumber\\
r_0 &=& \sqrt{X_0^2 + Y_0^2 + D^2} \, , \nonumber\\
\theta_0 &=& \arccos \left( \frac{Y_0 \sin i + D \cos i}{r_0} \right) \, , \nonumber\\
\phi_0 &=& \arctan \left( \frac{X_0}{D \sin i - Y_0 \cos i} \right) \, .
\ee
The initial 4-momentum of the photon is $k_0^\mu = \frac{\partial x^\mu}{\partial X^\mu} K_0^\mu$, where $\{ x^\mu \} = (t,x,y,z)$, $\{ X^\mu \} = (t,X,Y,Z)$ are the Cartesian coordinates of the distant observer, and $K_0^\mu = ( K_0 , 0 , 0 , - K_0)$ is the initial 4-momentum of the photon in the reference frame of the distant observer. The result is 
\be\label{eq-ks-initial}
k_0^r &=& - \frac{D}{r_0} \, K_0 \, , \nonumber\\
k_0^\theta &=& \frac{1}{ \sqrt{X_0^2 + \left( D \sin i - Y_0 \cos i\right)^2} } 
\left[ \cos i - \left( Y_0 \sin i + D \cos i\right) \frac{D}{r^2_0} \right] K_0 \, , \nonumber\\
k_0^\phi &=& \frac{X_0 \sin i}{ X_0^2 + \left( D \sin i - Y_0 \cos i\right)^2 } \, K_0 \, .
\ee
$k_0^t$ can be easily obtained from the condition $g_{\mu\nu} k_0^\mu k_0^\nu = 0$ with the metric tensor of flat spacetime 
\be\label{eq-kt-initial}
k_0^t &=& \sqrt{\left( k_0^r \right)^2 + r^2_0 \left( k_0^\theta \right)^2 + r_0^2 \sin^2\theta_0 \left( k_0^\phi \right)^2} \, .
\ee

With the initial conditions in Eq.~(\ref{eq-trtp-initial}), Eq.~(\ref{eq-ks-initial}), and Eq.~(\ref{eq-kt-initial}), we can integrate the geodesic equations in the spherical-like coordinates $(t,r,\theta,\phi)$ backwards in time from a point in the image plane of the distant observer to its emission point on the accretion disk. In the Kerr metric in Boyer-Lindquist coordinates, the equations of motion are separable and we can restrict the attention to the motion in the $r\theta$-plane; the corresponding equations can be solved in terms of elliptic integrals~\cite{Bambi:2017khi}. In general, this is not the case. We can still exploit the fact that the spacetime is stationary and axisymmetric, so we have Eq.~(\ref{eq-tdot}) and Eq.~(\ref{eq-phidot}), which can be rewritten as
\be
\frac{dt}{d\lambda'} &=& \frac{g_{\phi\phi} + b \, g_{t\phi}}{g_{t\phi}^2 - g_{tt} g_{\phi\phi}} \, , \\
\frac{d\phi}{d\lambda'} &=& - \frac{g_{t\phi} + b \, g_{tt}}{g_{t\phi}^2 - g_{tt} g_{\phi\phi}} \, ,
\ee 
where $b = L_z/E = - k_\phi/k_t$ is a constant along the photon trajectory and $\lambda' = E \lambda$ is the normalized affine parameter. For the $r$- and $\theta$-coordinates, we have to solve the second order geodesic equations, which reduces to the following equations for the metric in Eq.~(\ref{eq-metric})
\be
\frac{d^2r}{{d\lambda'}^2} &=& - \Gamma^r_{tt} \left( \frac{dt}{d\lambda'} \right)^2
- \Gamma^r_{rr} \left( \frac{dr}{d\lambda'} \right)^2
- \Gamma^r_{\theta\theta} \left( \frac{d\theta}{d\lambda'} \right)^2
- \Gamma^r_{\phi\phi} \left( \frac{d\phi}{d\lambda'} \right)^2
- 2 \Gamma^r_{t\phi} \left( \frac{dt}{d\lambda'} \right) \left( \frac{d\phi}{d\lambda'} \right)
- 2 \Gamma^r_{r\theta} \left( \frac{dr}{d\lambda'} \right) \left( \frac{d\theta}{d\lambda'} \right) \, , \qquad \\
\frac{d^2\theta}{{d\lambda'}^2} &=& - \Gamma^\theta_{tt} \left( \frac{dt}{d\lambda'} \right)^2
- \Gamma^\theta_{rr} \left( \frac{dr}{d\lambda'} \right)^2
- \Gamma^\theta_{\theta\theta} \left( \frac{d\theta}{d\lambda'} \right)^2
- \Gamma^\theta_{\phi\phi} \left( \frac{d\phi}{d\lambda'} \right)^2
- 2 \Gamma^\theta_{t\phi} \left( \frac{dt}{d\lambda'} \right) \left( \frac{d\phi}{d\lambda'} \right)
- 2 \Gamma^\theta_{r\theta} \left( \frac{dr}{d\lambda'} \right) \left( \frac{d\theta}{d\lambda'} \right) \, ,
\ee
where $\Gamma^\mu_{\nu\rho}$ are the Christoffel symbols of the metric.

%%%%%%%%%%%%%%%

\subsection{Redshift factor and emission angle}

When the photon hits the accreting material on the equatorial plane, we have to calculate the redshift factor $g$ and the emission angle $\vartheta_{\rm e}$

If the photon hits the accreting material in the disk region ($r_{\rm e} > r_{\rm ISCO}$), the redshift factor is 
\be\label{eq-redshiftfactor}
g = \frac{E_{\rm o}}{E_{\rm e}} = \frac{ - u_{\rm o}^\mu k_\mu }{ - u_{\rm e}^\mu k_\mu } \, ,
\ee
where $u_{\rm o}^\mu = ( 1 , 0 , 0 , 0 )$ is the 4-velocity of the distant observer, $u_{\rm e}^\mu = u_{\rm e}^t ( 1 , 0 , 0 , \Omega_{\rm K} )$ is the 4-velocity of the Keplerian material in the accretion disk, and $k_\mu = (k_t , k_r , k_\theta , k_\phi)$ is the conjugate 4-momentum of the photon. $k_\mu$ should be evaluated at the detection point in the numerator and at the emission point in the denominator, but actually $k_t$ and $k_\phi$ are constants of motion and $k_r$ and $k_\theta$ do not play any role as the $r$- and $\theta$-components of the 4-velocities of the observer and of the material in the disk vanish. With $u_{\rm e}^t $ from Eq.~(\ref{eq-tdot2}), the redshift factor $g$ is
\be
g = \frac{\sqrt{ - g_{tt} - 2 g_{t\phi} \Omega_{\rm K} - g_{\phi\phi} \Omega_{\rm K}^2 }}{1 - b \, \Omega_{\rm K}}
\ee
where $b = - k_\phi/k_t$ is a constant of motion and can be evaluate, for example, from the initial conditions, so from the last expression in Eq.~(\ref{eq-ks-initial}) and from Eq.~(\ref{eq-kt-initial}).

To evaluate the emission angle in the rest-frame of the material in the disk, we need the normal to the disk surface in the rest-frame of the material in the disk, $n^\mu$. We have to consider the locally Minkowskian reference frame associated to the material in the disk (see Appendix~\ref{a1}) and $n^\mu = E^\mu_{(Z)}$, so
\be
n^\mu = \frac{1}{\sqrt{g_{\theta\theta}}}
\left(
\begin{array}{c}
0 \\
0 \\
1 \\ 
0 
\end{array}
\right) .
\ee
The emission angle $\vartheta_{\rm e}$ can be evaluated as
\be\label{eq-emangle}
\cos \vartheta_{\rm e} = \frac{n^\mu k_\mu}{u_{\rm e}^\nu k_\nu} 
= \frac{g}{\sqrt{g_{\theta\theta}}} \frac{k_\theta}{k_t}
= \frac{1}{\sqrt{g_{\theta\theta}}} 
\frac{\sqrt{ - g_{tt} - 2 g_{t\phi} \Omega_{\rm K} - g_{\phi\phi} \Omega_{\rm K}^2 }}{1 - b \, \Omega_{\rm K}} \frac{k_\theta}{k_t} \, ,
\ee
where $k_\theta$ is the $\theta$-component of the conjugate 4-momentum of the photon at the emission point in the disk.

If the photon hits the accreting material in the plunging region ($r_{\rm e} < r_{\rm ISCO}$), the redshift factor is still given by Eq.~(\ref{eq-redshiftfactor}), but now the 4-velocity $u_{\rm e}^\mu$ has three non-vanishing components $(u_{\rm e}^t , u_{\rm e}^r , u_{\rm e}^\phi)$. The redshift factor is
\be\label{eq-g-plunging}
g = \left( u_{\rm e}^t + u_{\rm e}^r \frac{k_r}{k_t} + u_{\rm e}^\phi \frac{k_\phi}{k_t} \right)^{-1} \, ,
\ee
where $u_{\rm e}^t$, $u_{\rm e}^r$, and $u_{\rm e}^\phi$ are give, respectively, by the expressions in Eq.~(\ref{eq-tdot-plunging}), Eq.~(\ref{eq-rdot-plunging}), and Eq.~(\ref{eq-phidot-plunging}) if the material is in free fall. The locally Minkowskian reference frame associated to the material in the plunging region is different from that reported in Appendix~\ref{a1}, because $u_{\rm e}^r$ is non-vanishing in the plunging region and $E^\mu_{(T)} = u^\mu$. However, since $u_{\rm e}^\theta$ is still vanishing, we still have
\be
n^\mu = \frac{1}{\sqrt{g_{\theta\theta}}}
\left(
\begin{array}{c}
0 \\
0 \\
1 \\ 
0 
\end{array}
\right) ,
\ee
and the emission angle is
\be
\cos \vartheta_{\rm e} = \frac{g}{\sqrt{g_{\theta\theta}}} \frac{k_\theta}{k_t} 
= \frac{1}{\sqrt{g_{\theta\theta}}} 
\left( u_{\rm e}^t + u_{\rm e}^r \frac{k_r}{k_t} + u_{\rm e}^\phi \frac{k_\phi}{k_t} \right)^{-1} \frac{k_\theta}{k_t} \, .
\ee

%%%%%%%%%%%%%%%%%%%%%%%%%%%%%%%

\section{Novikov-Thorne Disk}

To calculate relativistic reflection spectra, the disk model discussed in Section~\ref{s-disk} is enough. We just assume that the disk is infinitesimally thin, perpendicular to the black hole spin axis, and that the material in the disk moves on geodesic, equatorial, circular orbits. There are no other assumptions. To calculate relativistic thermal spectra, these ingredients are not enough. We need a disk model predicting the temperature of the disk at any radial coordinate. The Novikov-Thorne disk model addresses this point~\cite{Novikov:1973kta,Page:1974he}. In this section, we consider a simplified version of the Novikov-Thorne model and we follow a more heuristic approach than the original publications, without paying attention to all assumptions behind. We consider an ideal case in which the system is in a steady-state configuration, while the actual model is valid even if the system is highly dynamical and derives the time-averaged properties of the disk. The details of the complete model can be found in Refs.~\cite{Novikov:1973kta,Page:1974he}.

%%%%%%%%%%%%%%%

\subsection{Near equatorial metric} 

The line element of our spacetime is given in Eq.~(\ref{eq-metric}) in spherical-like coordinates $(t,r,\theta,\phi)$. For the discussion of the Novikov-Thorne model, it is convenient to change coordinate system and choose cylindrical-like coordinates $(t,\rho,z,\phi)$. The coordinate transformation is
\be
\rho = r \sin\theta \, , \quad z = r \cos\theta \, , 
\ee
with inverse
\be
r = \sqrt{\rho^2 + z^2} \, , \quad \theta = \arctan \left( \frac{\rho}{z} \right) \, .
\ee
Since we are interested in the region near the equatorial plane, we expand the metric coefficients $g_{\rho\rho}$, $g_{zz}$, and $g_{\rho z}$ around $z=0$
\be
g_{\rho\rho} &=& \left[ 1 - \frac{z^2}{\rho^2} + O \left( z^4 \right) \right] g_{rr} 
+ \left[ \frac{z^2}{\rho^4} + O \left( z^4 \right) \right] g_{\theta\theta} = g_{rr} + O \left( z^2 \right) \, , \nonumber\\
g_{zz} &=& \left[ \frac{z^2}{\rho^2} + O \left( z^4 \right) \right] g_{rr}
+ \left[ \frac{1}{\rho^2} - \frac{2 z^2}{\rho^4} + O \left( z^4 \right) \right] g_{\theta\theta} 
= \frac{g_{\theta\theta} }{\rho^2} + O \left( z^2 \right) \, , \nonumber\\
g_{\rho z} &=& \left[ \frac{z}{\rho} + O \left( z^3 \right) \right] g_{rr}
+ \left[ - \frac{z}{\rho^3} + O \left( z^3 \right) \right] g_{\theta\theta} = O \left( z \right) \, .
\ee
If we ignore corrections of order $z/\rho$, the new line element is
\be
ds^2 = g_{tt} dt^2 + 2 g_{t\phi} dt d\phi + g_{rr} d\rho^2 + \frac{g_{\theta\theta}}{\rho^2} dz^2 + g_{\phi\phi} d\phi^2 \, .
\ee
In many black hole spacetimes (including the Kerr case), $g_{\theta\theta}/\rho^2 = 1$ on the equatorial plane. If this is not the case, we can always introduce the coordinate $dz' = \sqrt{g_{\theta\theta}/\rho^2} \, dz$ to have $g_{z' z'} = 1$. In the end, the line element in cylindrical-like coordinates $(t,r,z,\phi)$ of our spacetime near the equatorial plane can always be written as
\be\label{eq-metric-eqpl}
ds^2 &=& \tilde{g}_{tt} dt^2 + 2 \tilde{g}_{t\phi} dt d\phi + \tilde{g}_{rr} dr^2 + dz^2 + \tilde{g}_{\phi\phi} d\phi^2 \, .
\ee
where the tilde $\tilde{}$ is used to indicate the near equatorial metric coefficients in cylindrical-like coordinates $(t,r,z,\phi)$ and avoid confusion with the metric coefficients in spherical-like coordinates $(t,r,\theta,\phi)$ in Eq.~(\ref{eq-metric}). The infinitesimal element of the 4-volume of the metric in Eq.~(\ref{eq-metric-eqpl}) is
\be
dV = \sqrt{ - \tilde{g}} \, dt \, dr \, dz \, d\phi 
= \sqrt{ \tilde{g}_{rr} \left( \tilde{g}_{t\phi}^2 - \tilde{g}_{tt} \tilde{g}_{\phi\phi}\right) } \, dt \, dr \, dz \, d\phi \, .
\ee

%%%%%%%%%%%%%%%

\subsection{Radial structure of the disk}

From the laws of conservation of rest-mass, $\nabla_\mu \left( \rho u^\mu \right) = 0$, conservation of energy, $\nabla_\mu T^{t\mu} = 0$, and conservation of angular momentum, $\nabla_\mu T^{\phi\mu} = 0$, we can derive the radial structure of the accretion disk~\cite{Page:1974he}.

\begin{figure}[b]
\centering
\includegraphics[width=0.6\linewidth]{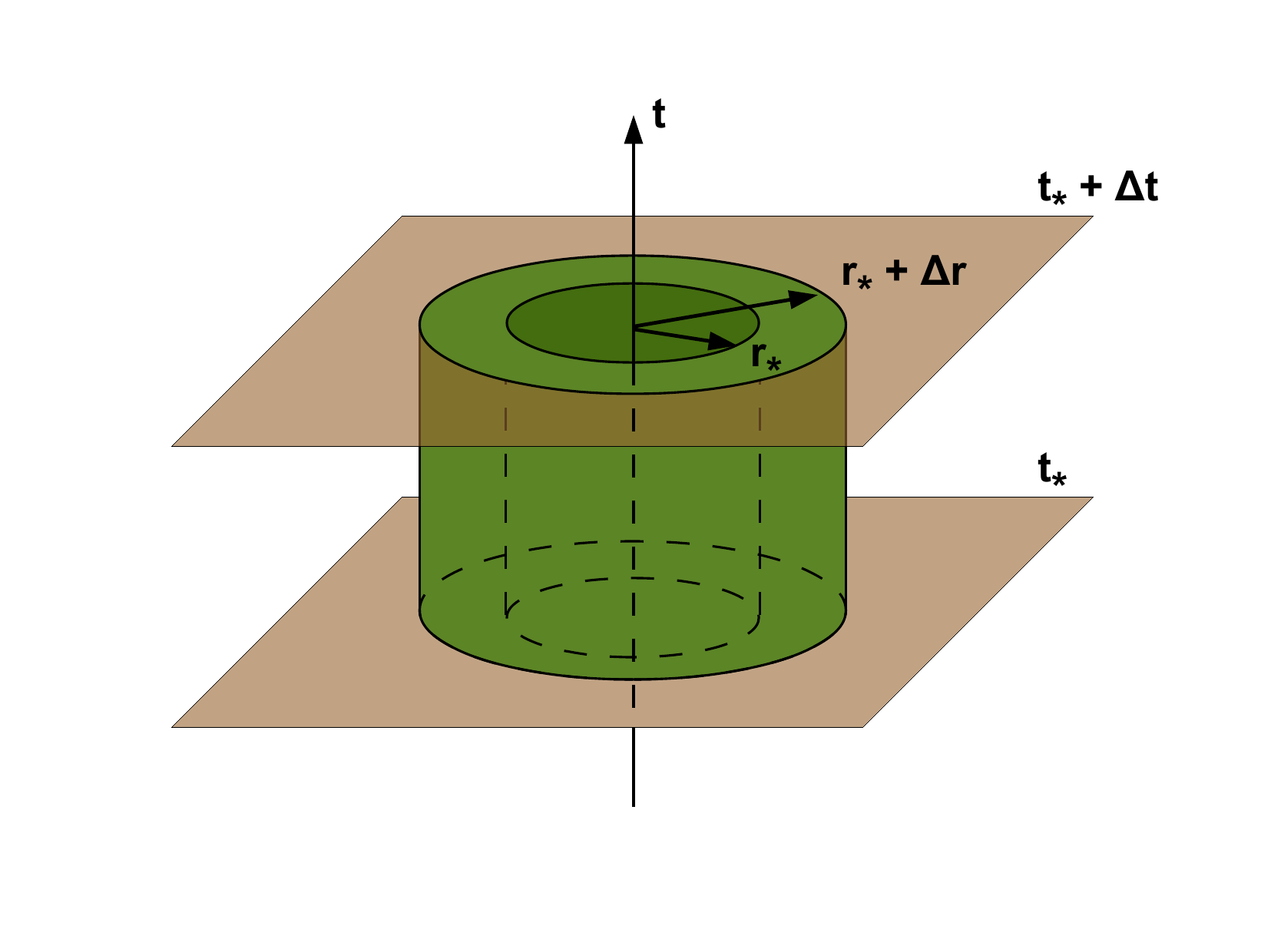}
\vspace{-1.2cm}
\caption{The 4-volume $V$ in Eq.~(\ref{eq-volume}) is the cylinder-like region of the accretion disk between the surfaces $r = r_*$ and $r = r_* + \Delta r$ and the surfaces $t = t_*$ and $t = t_* + \Delta t$.}
\label{f-mdot}
\end{figure}

Let us consider the conservation of rest-mass, $\nabla_\mu \left( \rho u^\mu \right) = 0$, where $\rho$ is the mass density measured in the rest-frame of the material ($\rho = m n$, where $m$ is the mean rest-mass per particle and $n$ is the particle density in the rest-frame of the material). We integrate this expression over the 4-volume of the spacetime from $t_*$ to $t_* + \Delta t$ and from $r_*$ to $r_* + \Delta r$ shown in Fig.~\ref{f-mdot}, where we collapsed the $z$ direction  
\be\label{eq-volume}
0 = \int_{V} \nabla_\mu \left( \rho u^\mu \right) \, \sqrt{ - \tilde{g}} \, dt \, dr \, dz \, d\phi 
= \int_{V} \frac{\partial}{\partial x^\mu} \left( \sqrt{ - \tilde{g}} \rho u^\mu \right) \, dt \, dr \, dz \, d\phi
= \int_{\Sigma} \sqrt{ - \tilde{g}} \rho u^\mu \, d\sigma_\mu \, ,
\ee
where first we have used the formula for the covariant divergence of a generic vector $A^\mu$ in a generic spacetime with metric $g_{\mu\nu}$ and metric determinant $g$ (see, for instance, Section~5.3 in Ref.~\cite{Bambi:2018drb} for its derivation)
\be
\nabla_\mu A^\mu = \frac{1}{\sqrt{ - g}} \frac{\partial}{\partial x^\mu} \left( \sqrt{ - g} A^\mu \right) \, ,
\ee
and then we have applied Gauss's theorem to convert the integral over the volume $V$ to the integral over its surface $\Sigma$. As shown in Fig.~\ref{f-mdot}, in our case we can consider four surfaces: two cylindrical-like surfaces at, respectively, $r = r_*$ and $r = r_* + \Delta r$, and two annulus-like surface at, respectively, $t = t_*$ and $t = t_* + \Delta t$. We have thus four integrals 
\be\label{eq-integral123}
0 &=& \left[ \int_{t_*}^{t_* + \Delta t} dt \int_{-h/2}^{h/2} dz \int_{0}^{2\pi} d\phi \, \sqrt{ - \tilde{g}} \rho u^r \right]_{r = r_* + \Delta r}
- \left[ \int_{t_*}^{t_* + \Delta t} dt \int_{-h/2}^{h/2} dz \int_{0}^{2\pi} d\phi \, \sqrt{ - \tilde{g}} \rho u^r \right]_{r = r_*} \nonumber\\
&& + \left[ \int_{r_*}^{r_* + \Delta r} dr \int_{-h/2}^{h/2} dz \int_{0}^{2\pi} d\phi \, \sqrt{ - \tilde{g}} \rho u^t \right]_{t = t_* + \Delta t} 
- \left[ \int_{r_*}^{r_* + \Delta r} dr \int_{-h/2}^{h/2} dz \int_{0}^{2\pi} d\phi \, \sqrt{ - \tilde{g}} \rho u^t \right]_{t = t_*} \, .
\ee
where $h$ is the thickness of our disk and, for simplicity, we can assume that the rest-mass density is $\rho = {\rm constant}$ for $-h/2 < z < h/2$ and 0 otherwise. In our simple model in steady-state, the last two integral exactly cancel each other because the integrand is independent of time. The rest-mass accretion rate is
\be\label{eq-def-massdot} 
\dot{M} = \int_{-h/2}^{h/2} dz \int_{0}^{2\pi} d\phi \, \sqrt{ - \tilde{g}} \rho u^r = 2 \pi \sqrt{ - \tilde{g}} \rho h u^r \, ,
\ee
and therefore Eq.~(\ref{eq-integral123}) becomes
\be
\Delta t \, \dot{M} \left( r_* + \Delta r \right) - \Delta t \, \dot{M} \left( r_* \right) = 0 \, ,
\ee
so the rest-mass accretion rate (\ref{eq-def-massdot}) is independent of $r$. This could have been expected, because there are no disk winds or outflows in our model.

One can proceed in a similar way with the conservation of energy and angular momentum to find the energy flux from the surface of the accretion disk and the torque at every radial coordinate $r$. The details can be found in Ref.~\cite{Page:1974he}. To calculate the relativistic thermal spectrum of a disk, we need to have the energy flux from the surface of the disk, which is given by the following expression
\be\label{eq-time-averaged-flux}
\mathcal{F} (r) = \frac{\dot{M}}{4 \pi M^2} \, \hat{F} (r) \, ,
\ee
where $\hat{F}$ is the following dimensionless function
\be\label{eq-f-hat}
\hat{F} (r) = - \frac{1}{\left( E - \Omega_{\rm K} L_z \right)^2} \frac{\partial \Omega_{\rm K}}{\partial r}
\frac{M^2}{\sqrt{- \tilde{g}}} \int_{r_{\rm in}}^{r} \left( E - \Omega_{\rm K} L_z \right) \,
\frac{\partial L_z}{\partial \varrho} \, d\varrho \, .
\ee
In Eq.~(\ref{eq-f-hat}), $E$ and $L_z$ are, respectively, the specific energy and the axial component of the specific angular momentum for geodesic, equatorial, circular orbits given in Eq.~(\ref{eq-E-c-orbit}) and Eq.~(\ref{eq-L-c-orbit}), $\Omega_{\rm K}$ is the Keplerian angular velocity of the material in the disk given in Eq.~(\ref{eq-omega}), and $r_{\rm in}$ is the inner edge of the accretion disk ($r_{\rm in} \ge r_{\rm ISCO}$).

%%%%%%%%%%%%%%%

\subsection{Plunging region}

From Eq.~(\ref{eq-def-massdot}) we can estimate the rest-mass density in the plunging region. If we assume that the radial component of the 4-velocity of the accreting material is given by Eq.~(\ref{eq-rdot-plunging}) in the plunging region and $h$ is roughly constant (say $h \sim 0.1$ or 0.01~$M$), the rest-mass density is
\be\label{eq-density-plunging}
\rho (r) = \frac{\dot{M}}{2 \pi \, h } 
\Big[ \tilde{g}_{\phi\phi} (E^{\rm ISCO})^2 + 2 \tilde{g}_{t\phi} 
E^{\rm ISCO} L^{\rm ISCO}_z + \tilde{g}_{tt} (L_z^{\rm ISCO})^2 - \tilde{g}_{t\phi}^2 + \tilde{g}_{tt} \tilde{g}_{\phi\phi} \Big]^{-1/2} \, .
\ee
For a fixed background metric, $\rho$ in Eq.~(\ref{eq-density-plunging}) is only a function of $r$ and $\dot{M}$.

%%%%%%%%%%%%%%%%%%%%%%%%%%%%%%%

\section{Relativistic Thermal Spectrum}

\begin{figure}[b]
\centering
\includegraphics[width=0.5\linewidth]{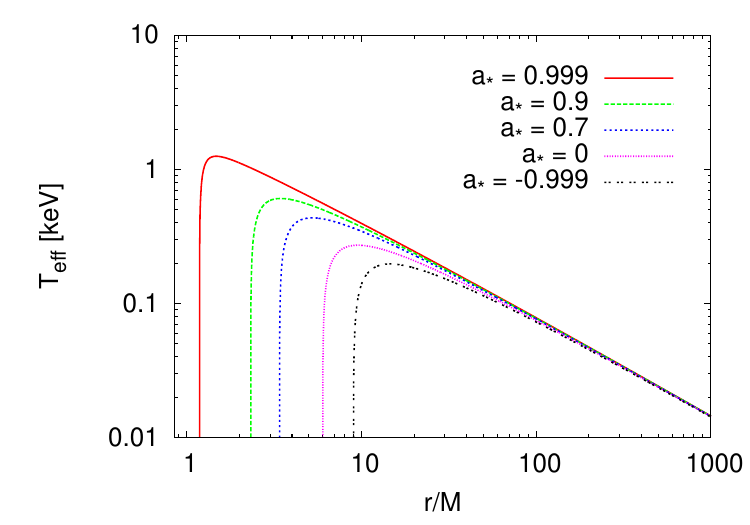}
\vspace{-0.3cm}
\caption{Radial profile of the effective temperature $T_{\rm eff}$ in Novikov-Thorne disks around Kerr black holes with $M = 10$~$M_\odot$, $\dot{M} = 10^{18}$~g~s$^{-1}$, and different values of the black hole spin parameters $a_*$.}
\label{f-tprofile}
\end{figure}

If we assume that $i)$ the material in the accretion disk is in local thermal equilibrium and $ii)$ the heat transport along the radial direction is negligible and energy and angular momentum are radiated from the disk surface, we can define an effective temperature $T_{\rm eff}$ at every radial coordinate by imposing that the energy flux $\mathcal{F}$ in Eq.~(\ref{eq-time-averaged-flux}) is the heat power emitted from the surface of the disk and using the Stefan-Boltzmann Law
\be
\mathcal{F} = \sigma_{\rm SB} \, T^4_{\rm eff} \, ,
\ee
where $\sigma_{\rm SB} = 5.67 \cdot 10^{-5}$~erg~s$^{-1}$~cm$^{-2}$~K$^{-4}$ is the Stefan-Boltzmann constant. Fig.~\ref{f-tprofile} shows the radial profile of the effective temperature of a Novikov-Thorne disk around a Kerr black hole with mass $M = 10$~$M_\odot$ and different values of the black hole spin parameters $a_*$ assuming that the black hole mass accretion rate is $\dot{M} = 10^{18}$~g~s$^{-1}$ and that $r_{\rm in} = r_{\rm ISCO}$. Since Novikov-Thorne disks are realized when the black hole mass accretion rate is of the order of 10\% of the Eddington limit of the source, $\mathcal{F}$ turns out to be always proportional to $1/M$, and we can easily see that the effective temperature of the inner part of a Novikov-Thorne disk should be of the order of 1~keV for $M = 10$~$M_\odot$ and of order of 10~eV for $M = 10^9$~$M_\odot$
\be
T_{\rm eff} \sim 1 \left( \frac{10 \, M_\odot}{M} \right)^{1/4} {\rm keV} \, .
\ee

In the case of accretion disks around stellar-mass black holes, the effective temperature of the inner part is high and non-thermal effects (mainly electron scattering in the disk atmosphere) cannot be ignored. Deviations from the blackbody spectrum can be taken into account by introducing the {\it color factor} (or hardening factor) $f_{\rm col}$ and defining the {\it color temperature} $T_{\rm col} = f_{\rm col} T_{\rm eff}$. The specific intensity of the radiation in the rest-frame of the material in the disk is (for this formula it can be useful to reintroduce fundamental constants rather than using natural units) 
\be\label{eq-si-thermal}
I_{\rm e} = \frac{2 h \nu_{\rm e}^3}{c^2} \frac{1}{f_{\rm col}^4} 
\frac{\Upsilon}{\exp\left(\frac{h \nu_{\rm e}}{k_{\rm B} T_{\rm col}}\right) - 1} \, ,
\ee
where $h$ is the Planck constant, $c$ is the speed of light, $k_{\rm B}$ is the Boltzmann constant, and $\Upsilon = \Upsilon ( \vartheta_{\rm e} )$ is a function regulating the angular emission. The two most popular choices are $\Upsilon = 1$ (isotropic emission) and $\Upsilon = 0.5 + 0.75 \cos \vartheta_{\rm e}$ (limb-darkened emission). In the case of a 10~$M_\odot$ black hole with an accretion luminosity of 10\% its Eddington limit, $f_{\rm col}$ is expected to be in the range 1.5 to 1.9 and it can be evaluated by a model for the disk atmosphere~\cite{Shimura:1995nu,Davis:2006bk,Davis:2004jf}. Note that $I_{\rm e} = I_{\rm e} (r_{\rm e})$ because $T_{\rm col}$ (and at some level even $f_{\rm col}$) depends on the radial coordinate.

With $I_{\rm e}$ in Eq.~(\ref{eq-si-thermal}), we can proceed as in the case of the calculation of a relativistic reflection spectrum discussed in Section~\ref{s-relref} and infer the relativistic thermal spectrum of a source through Eq.~(\ref{eq-oflux}). We consider a distant observer and we fire photons from the plane of the distant observer to the accretion disk with the photon initial conditions presented in Subsection~\ref{ss-initialcond}. When a photon hits the accretion disk ($r > r_{\rm ISCO}$), we calculate the redshift factor $g$ and, if we do not assume $\Upsilon = 1$, the emission angle $\vartheta_{\rm e}$. We repeat the calculations for every small element $dXdY$ in the image plane of the distant observer and then we integrate over the full image to get the relativistic thermal spectrum of the whole disk. In the Novikov-Thorne model, there is no emission of radiation from the plunging region. However, there is material even there and one can include the thermal spectrum of the plunging region; see, for instance, the model in Ref.~\cite{Mummery:2024mrq}.

%%%%%%%%%%%%%%%%%%%%%%%%%%%%%%%

\section{Cunningham's Transfer Function}

The calculations of relativistic reflection spectra and of relativistic thermal spectra of thin disks discussed in the previous sections turn out to be very time-consuming. The geodesic equations are relatively easy and fast to solve, but normally one has to calculate the trajectories of millions of photons to reach the required accuracy in the final result. The calculation of non-relativistic reflection spectra requires to solve radiative transfer equations, which are definitively time-consuming. On the other hand, during the data analysis process we have to be able to generate quickly many spectra for different values of the model parameters in order to scan the full parameter space and find the best-fit model, so we cannot do the calculations described in the previous sections.

Every point in the image of the accretion disk in the plane of the distant observer is specified by the Cartesian coordinates $(X,Y)$. After calculating the photon trajectories, for every point in the image of the accretion disk we can associate a point on the accretion disk and we can determine its emission radius $r_{\rm e}$ and its redshift factor $g$. In the case of an infinitesimally thin, Keplerian accretion disk perpendicular to the spin axis of a Kerr black hole, every point on the accretion disk is visible to a distant observer: the circle of the points with the same emission radius $r_{\rm e}$ will not be a circle in the image of the accretion disk in the plane of the distant observer but it will be still a closed loop. Moreover, for each of these closed loops, we will have a point with the minimum value of the redshift factor and a point with the maximum value of the redshift factor. Which points on the accretion disk have the minimum and maximum redshift factors depends on the emission radius $r_{\rm e}$ and the inclination angle of the disk $i$ (if we fix the background metric by specifying the black hole spin parameter $a_*$) because the redshift factor is the result of the competition/combination between the Doppler boosting (which depends on both the emission radius and the inclination angle of the disk) and the gravitational redshift (which depends only on the emission radius). We can thus write $g_{\rm min} = g_{\rm min} (r_{\rm e} , i)$ and $g_{\rm max} = g_{\rm max} (r_{\rm e} , i)$ to indicate the minimum and maximum redshift factor for the emission radius $r_{\rm e}$ and the disk inclination angle $i$. Since the points on the accretion disk at the same emission radius form a closed loop on the image of the disk in the plane of the observer, the point with minimum redshift factor and the point with maximum redshift factor are connected by two branches. It turns out that, in both branches, the redshift factor monotonically increases when we move from the point with minimum redshift factor to the point with maximum redshift factor\footnote{These results, strictly speaking, are valid in the case of infinitesimally thin, Keplerian accretion disk in Kerr spacetimes. They are normally valid even in the case of infinitesimally thin, Keplerian accretion disk in non-Kerr spacetimes, but exceptions are possible. In the case of accretion disks of finite thickness, some parts of the disk may not be visible to a distant observer, so the image of the points on the accretion disk at the same emission radius may not be a closed loop~\cite{Abdikamalov:2020oci}.}. These results allow us to define the relative redshift factor $g^*$ as
\be
g^* = \frac{g - g_{\rm min}}{g_{\rm max} - g_{\rm min}} \, ,
\ee
and to parametrize the points of the accretion disk with the emission radius $r_{\rm e}$ and the relative redshift factor $g^*$.

\begin{figure}[b]
\centering
\includegraphics[width=0.45\linewidth]{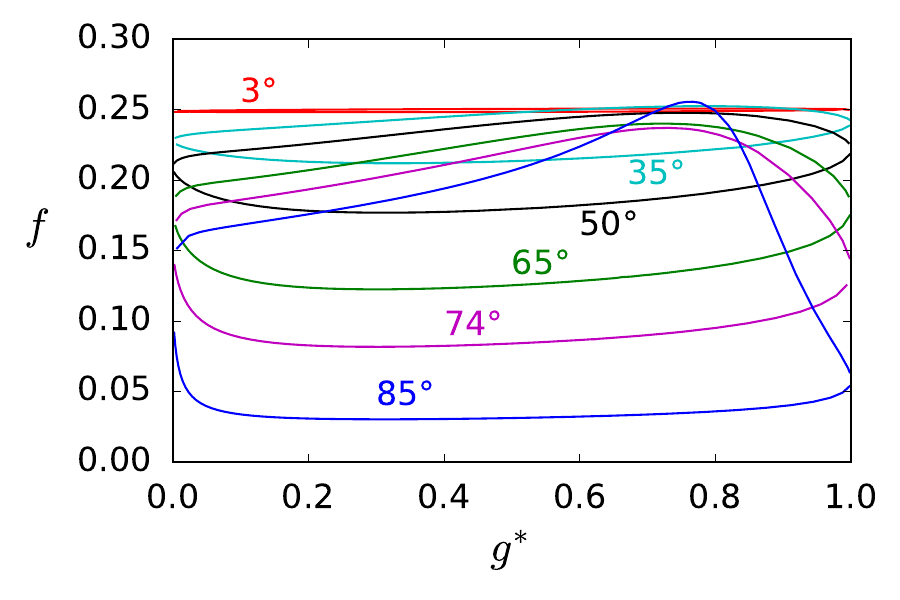}
\hspace{0.5cm}
\includegraphics[width=0.45\linewidth]{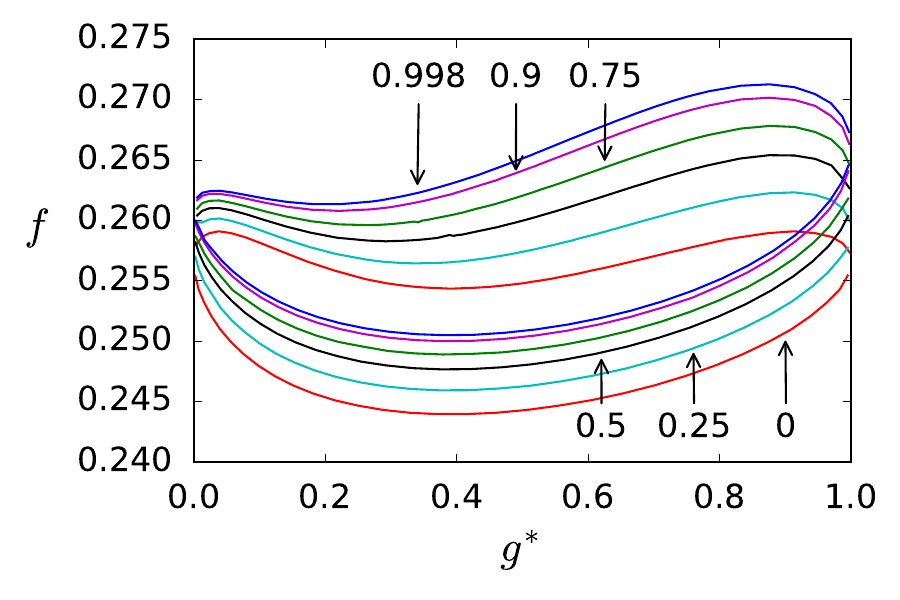}
\vspace{-0.4cm}
\caption{Examples of Cunningham's transfer function for an infinitesimally thin, Keplerian accretion disk in the Kerr spacetime. Left panel: the central black hole has spin parameter $a_* = 0.998$ and the plot shows the transfer functions at the emission radius $r_{\rm e} = 4$~$M$ for different values of the inclination angle of the disk $i$. Right panel: the emission radius is $r_{\rm e} = 7$~$M$ and the inclination angle of the disk is $i = 30^\circ$ and the plot shows the transfer functions for different values of the black hole spin parameter $a_*$. Figures from Ref.~\cite{Bambi:2016sac}.}
\label{f-trf}
\end{figure}

We can thus recast Eq.~(\ref{eq-oflux}) in the following form~\cite{Cunningham:1975zz,Speith:1995xxx,Bambi:2016sac}
\be\label{eq-2branches}
F_{\rm o} (\nu_{\rm o}) &=& \frac{1}{D^2} \int_{r_{\rm in}}^{r_{\rm out}} dr_{\rm e} \int_0^1 dg^* \, 
\frac{\pi r_{\rm e} g^2}{\sqrt{g^* \left( 1 - g^* \right)}} 
\, f^{(1)} ( g^* , r_{\rm e} , i ) \, I_{\rm e} (\nu_{\rm e} , r_{\rm e} , \vartheta^{(1)}_{\rm e}) \nonumber\\
&& + \frac{1}{D^2} \int_{r_{\rm in}}^{r_{\rm out}} dr_{\rm e} \int_0^1 dg^* \, 
\frac{\pi r_{\rm e} g^2}{\sqrt{g^* \left( 1 - g^* \right)}} 
\, f^{(2)} ( g^* , r_{\rm e} , i ) \, I_{\rm e} (\nu_{\rm e} , r_{\rm e} , \vartheta^{(2)}_{\rm e}) \, ,
\ee
where $f$ is {\it Cunningham's transfer function}~\cite{Cunningham:1975zz}
\be
f^{(i)} ( g^* , r_{\rm e} , i ) = \frac{g \sqrt{g^* \left( 1 - g^* \right)}}{\pi r_{\rm e}} \left| \frac{\partial \left( X , Y \right)}{\partial \left( r_{\rm e} , g^* \right)} \right| \, 
\ee
and at every emission radius we have two transfer functions, $f^{(1)}$ and $f^{(2)}$, because we have two branches connecting $g_{\rm min}$ to $g_{\rm max}$. $r_{\rm in}$ and $r_{\rm out}$ are, respectively, the inner and the other radius of the accretion disk. $\left| \partial \left( X , Y \right) / \partial \left( r_{\rm e} , g^* \right) \right| $ is the Jacobian between the Cartesian coordinate of the image plane of the distant observer, $(X,Y)$, and the coordinates used on the accretion disk, $(r_{\rm e} , g^*)$. If the specific intensity at the emission point does not depend on the emission radius, $I_{\rm e} = I_{\rm e} (\nu_{\rm e} , r_{\rm e})$, and we can rewrite Eq.~(\ref{eq-2branches}) as
\be\label{eq-2branchesbis}
F_{\rm o} (\nu_{\rm o}) = \frac{1}{D^2} \int_{r_{\rm in}}^{r_{\rm out}} dr_{\rm e} \int_0^1 dg^* \, 
\frac{\pi r_{\rm e} g^2}{\sqrt{g^* \left( 1 - g^* \right)}} 
\left[ f^{(1)} ( g^* , r_{\rm e} , i ) + f^{(2)} ( g^* , r_{\rm e} , i ) \right] \, I_{\rm e} (\nu_{\rm e} , r_{\rm e}) \, .
\ee
Fig.~\ref{f-trf} shows some examples of transfer function of an infinitesimally thin, Keplerian accretion disk in the Kerr spacetime.

In the case of a relativistic thermal model, the specific intensity at the emission point is given by Eq.~(\ref{eq-si-thermal}). The transfer function $f$ can be pre-calculated and tabulated into a FITS file\footnote{FITS (Flexible Image Transport System) is a common format for astronomical data files that can store multidimensional arrays and tables.} before the data analysis process. The model for data analysis has simply to solve the integral in Eq.~(\ref{eq-2branches}), or in Eq.~(\ref{eq-2branchesbis}) if $\Upsilon = 1$, and call the FITS file to know the value of the transfer function. This is how the model {\tt nkbb} works~\cite{Zhou:2019fcg}.

For a relativistic reflection model, there are a few complications. The specific intensity at the emission point can be determined after solving numerically radiative transfer equations rather than from a simple analytical formula as in the case of the thermal spectrum. The emissivity profile of the accretion disk can be determined after studying how the corona illuminates the disk. Unlike in the case of the thermal component, we do not know the actual luminosity of the disk in terms of parameters like the black hole mass and the mass accretion rate, and therefore we cannot predict the observed flux as a function of the distance of the source. It is thus convenient to rewrite Eq.~(\ref{eq-2branches}) as
\be\label{eq-obflux-refl}
F_{\rm o} (\nu_{\rm o}) &=& C \int_{r_{\rm in}}^{r_{\rm out}} dr_{\rm e} \int_0^1 dg^* \, 
\frac{\pi r_{\rm e} g^2}{\sqrt{g^* \left( 1 - g^* \right)}} 
\, f^{(1)} ( g^* , r_{\rm e} , i ) \, I_{\rm e} (F_X, \nu_{\rm e} , r_{\rm e} , \vartheta^{(1)}_{\rm e}) \nonumber\\
&& + C \int_{r_{\rm in}}^{r_{\rm out}} dr_{\rm e} \int_0^1 dg^* \, 
\frac{\pi r_{\rm e} g^2}{\sqrt{g^* \left( 1 - g^* \right)}} 
\, f^{(2)} ( g^* , r_{\rm e} , i ) \, I_{\rm e} (F_X, \nu_{\rm e} , r_{\rm e} , \vartheta^{(2)}_{\rm e}) \, ,
\ee
where $C$ is a normalization constant to be determined when we fit the data and $F_X = F_X (r_{\rm e})$ is the X-ray spectral flux illuminating the disk at the emission radius $r_{\rm e}$. $F_X$ is determined when we study how the corona and the returning radiation illuminate the disk.

Current reflection models for data analysis often employ the simplification to assume the same non-relativistic reflection spectrum over the whole disk and normalize the specific intensity through the emissivity profile $\epsilon (r_{\rm e})$. Within such an approximation, Eq.~(\ref{eq-obflux-refl}) can be written as  
\be
F_{\rm o} (\nu_{\rm o}) &=& C \int_{r_{\rm in}}^{r_{\rm out}} dr_{\rm e} \int_0^1 dg^* \, 
\frac{\pi r_{\rm e} g^2}{\sqrt{g^* \left( 1 - g^* \right)}} 
\left[ f^{(1)} ( g^* , r_{\rm e} , i ) + f^{(2)} ( g^* , r_{\rm e} , i ) \right]
\epsilon (r_{\rm e}) \, \bar{I}_{\rm e} (\nu_{\rm e}) \, ,
\ee
where $\bar{I}_{\rm e}$ is the ``average'' specific intensity over the disk (or some region of the disk if we divide the disk in a few zones with different specific intensity). It is evaluated from the weighted sum of specific intensities on the disk (a certain zone of the disk) with different emission angles\footnote{This approach is motivated by the need to be able to calculate quickly a relativistic reflection spectrum. Note that the ionization parameter is $\xi (r) = 4 \pi F / n$, where $F$ is the total flux illuminating the disk at the radial coordinate $r$ and $n$ is the electron density of the disk at the same radial coordinate. A model in which the ionization parameter $\xi$ and the disk electron density $n$ are constant over the disk would require that $F$ is constant too. This is not how most of the current models work: they employ $\bar{I}_{\rm e}$, which is calculated for certain values of $\xi$ and $n$, and model the emissivity profile with $\epsilon (r_{\rm e})$.}. Now the transfer function $f$, the emissivity profile $\epsilon$, and the specific intensity $\bar{I}_{\rm e}$ can be pre-calculated and tabulated into three different FITS files before the data analysis process. In the Kerr spacetime, the transfer function $f$ depends on the black hole spin parameter $a_*$ and the inclination angle of the disk $i$ and, once these two parameters are fixed, it is a function of $r_{\rm e}$ and $g^*$. The emissivity profile $\epsilon$ depends on the parameters of the specific coronal model (for example, in the lamppost setup it depends only on the height of the corona) and on the background metric (the black hole spin parameters $a_*$ in the case of the Kerr metric): after fixing the coronal model and the background metric, $\epsilon$ only depends on the emission radius $r_{\rm e}$. $\bar{I}_{\rm e}$ depends on the X-ray spectral flux illuminating the disk and the parameters of the accretion disk model. For example, if we assume that the X-ray spectral flux illuminating the disk can be approximated by a power law with a high energy cutoff, we can have 5~parameters: the photon index $\Gamma$, the high-energy cutoff $E_{\rm cut}$, the electron density of the accretion disk $n_{\rm e}$, the ionization parameter $\xi$, and the iron abundance $A_{\rm Fe}$. During the data analysis process, the model calls the three FITS files in which $f$, $\epsilon$, and $\bar{I}_{\rm e}$ are tabulated and quickly calculates different relativistic spectra for different values of the model parameters.

%%%%%%%%%%%%%%%%%%%%%%%%%%%%%%%

\section{Concluding remarks}

These notes review the relativistic calculations of the electromagnetic spectrum of cold, thin, Keplerian accretion disks around black holes. The disk-corona model predicts a thermal spectrum from the disk, a Comptonized spectrum from the hot corona, and a relativistically blurred reflection spectrum from the disk. The expressions reported in these notes are valid for a generic stationary, axisymmetric, asymptotically flat, circular black hole spacetime, and therefore they can be potentially applied to a large class of black hole solutions. The presentation in these notes is intentionally pedagogical in order to help graduate students to enter this line of research. More details can be found in the papers in the reference list. As shown in these notes, current models for data analysis present many simplifications in order to be able to produce quickly many spectra. The next generation of X-ray missions (e.g., \textsl{eXTP}, \textsl{Athena}, \textsl{STROBE-X}, \textsl{HEX-P}, etc.) promise to provide unprecedented high-quality data and they will require more advanced models than those available today.

%%%%%%%%%%%%%%%%%%%%%%%%%%%%%%%

%\begin{acknowledgments}
%This work was supported by ...
%\end{acknowledgments}

%%%%%%%%%%%%%%%%%%%%%%%%%%%%%%%

\appendix

\section{Locally Minkowskian reference frames}\label{a1}

There are situations in which it is convenient to choose the reference frame of a locally Minkowskian observer. Formally, this is equivalent to a coordinate transformation from the coordinate system $\{ x^\mu \}$ to the coordinate system $\{ {x'}^\mu \} = (T,X,Y,Z)$\footnote{The transformation (\ref{eq-tran-lmo}) reduces the metric tensor $g_{\mu\nu}$ to the Minkowski metric at a point of the spacetime, which is always possible because it is equivalent to make diagonal a symmetric matrix with constant coefficients and then rescale the coordinates to reduce the diagonal elements to $\pm 1$. See, for example, Ref.~\cite{Bambi:2018drb} for more details.}
\be\label{eq-tran-lmo}
dx^\mu \rar {dx'}^{(\alpha)} = E^{(\alpha)}_\mu dx^\mu \, ,
\ee
such that the new metric tensor is the Minkowski metric
\be
g_{\mu\nu} \rar \eta_{(\alpha)(\beta)} = E^\mu_{(\alpha)} E^\nu_{(\beta)} g_{\mu\nu} \, ,
\ee
where $E^\mu_{(\alpha)}$ are the inverse of $E^{(\alpha)}_\mu$, so $E^{(\alpha)}_\mu E_{(\alpha)}^\nu = \delta_\mu^\nu$ and $E^{(\alpha)}_\mu E_{(\beta)}^\mu = \delta_{(\beta)}^{(\alpha)}$. $\{ E^\mu_{(\alpha)} \}$ is the {\it tetrad} of the orthogonal basis vectors associated to the reference frame of the locally Minkowskian observer. If a vector (dual vector) has components $V^\mu$ ($V_\mu$) in the coordinate system $\{ x^\mu \}$, the components of the vector (dual vector) in the locally Minkowskian reference frame are
\be
V^{(\alpha)} = E^{(\alpha)}_\mu V^\mu \, , \quad V_{(\alpha)} = E_{(\alpha)}^\mu V_\mu \, .
\ee
It is straightforward to see that
\be
V^\mu = E_{(\alpha)}^\mu V^{(\alpha)} \, , \quad V_\mu = E^{(\alpha)}_\mu V_{(\alpha)} \, .
\ee
Note that spacetime indices are lowered by $g_{\mu\nu}$ and raised by $g^{\mu\nu}$ and internal indices of the locally Minkowskian reference frame are lowered by $\eta_{(\alpha)(\beta)}$ and raised by $\eta^{(\alpha)(\beta)}$.

The time-like tetrad basis vector $E^\mu_{(T)}$ is the 4-velocity of the locally Minkowskian observer in the coordinate system $\{ x^\mu \}$, say $u^\mu$. In our case, the spacetime metric is given in Eq.~(\ref{eq-metric}) in spherical-like coordinates $(t,r,\theta,\phi)$. If the locally Minkowskian observer is in a circular orbit (not necessarily on the equatorial plane), his/her 4-velocity is $u^\mu = u^t \, ( 1 , 0 , 0 , \omega)$, where $u^t$ can be inferred from $g_{\mu\nu} \dot{x}^\mu \dot{x}^\nu = -1$
\be
u^t = \frac{1}{\sqrt{-g_{tt} - 2 g_{t\phi} \omega - g_{\phi\phi} \omega^2}}
\ee
and $\omega = u^\phi / u^t$ is the angular velocity of the observer with respect to the coordinate system $(t,r,\theta,\phi)$. A natural choice for the tetrad of the orthogonal basis vectors associated to the reference frame of the locally Minkowskian observer is
\be\label{eq-tetrad}
E^\mu_{(T)} = u^t 
\left(
\begin{array}{c}
1 \\
0 \\
0 \\ 
\omega 
\end{array}
\right) , \;\;
%%%
E^\mu_{(X)} = 
\frac{1}{\sqrt{g_{rr}}}
\left(
\begin{array}{c}
0 \\
1 \\
0 \\ 
0 
\end{array}
\right) , \;\;
%%%
E^\mu_{(Y)} = 
\frac{u^t}{\sqrt{g^2_{t\phi} - g_{tt} g_{\phi\phi} }}
\left(
\begin{array}{c}
g_{t\phi} + \omega g_{\phi\phi} \\
0 \\
0 \\ 
-g_{tt} - \omega g_{t\phi}
\end{array}
\right) , \;\; 
%%%
E^\mu_{(Z)} = 
\frac{1}{\sqrt{g_{\theta\theta}}}
\left(
\begin{array}{c}
0 \\
0 \\
1 \\ 
0 
\end{array}
\right) . \qquad
\ee
The inverse is
\be\label{eq-tetrad-inverse}
E^{(T)}_\mu = - u^t 
\left(
\begin{array}{c}
g_{tt} + \omega g_{t\phi} \\
0 \\
0 \\ 
g_{t\phi} + \omega g_{\phi\phi}
\end{array}
\right) , \;\;
%%%
E^{(X)}_\mu = 
\sqrt{g_{rr}}
\left(
\begin{array}{c}
0 \\
1 \\
0 \\ 
0 
\end{array}
\right) , \;\;
%%%
E^{(Y)}_\mu = 
u^t \sqrt{g^2_{t\phi} - g_{tt} g_{\phi\phi} }
\left(
\begin{array}{c}
- \omega \\
0 \\
0 \\ 
1
\end{array}
\right) , \;\; 
%%%
E^{(Z)}_\mu = 
\sqrt{g_{\theta\theta}}
\left(
\begin{array}{c}
0 \\
0 \\
1 \\ 
0 
\end{array}
\right) . \qquad
\ee

A special case is an observer with angular velocity $\omega = - g_{t\phi}/g_{\phi\phi}$, which is often referred to as a {\it locally non-rotating frame} (LNRF) in the literature~\cite{Bardeen:1972fi}. The 4-velocity of such an observer is
\be
u^t = \sqrt{\frac{g_{\phi\phi}}{ g_{t\phi}^2 - g_{tt} g_{\phi\phi}}} \, .
\ee 
The tetrad of the orthogonal basis vectors associated to the locally non-rotating observer reduces to
\be\label{eq-tetrad-lnro}
E^\mu_{(T)} = \sqrt{\frac{g_{\phi\phi}}{ g_{t\phi}^2 - g_{tt} g_{\phi\phi}}}
\left(
\begin{array}{c}
1 \\
0 \\
0 \\ 
\omega 
\end{array}
\right) , \;\;
%%%
E^\mu_{(X)} = 
\frac{1}{\sqrt{g_{rr}}}
\left(
\begin{array}{c}
0 \\
1 \\
0 \\ 
0 
\end{array}
\right) , \;\;
%%%
E^\mu_{(Y)} = 
\frac{1}{\sqrt{g_{\phi\phi}}}
\left(
\begin{array}{c}
0 \\
0 \\
0 \\ 
1
\end{array}
\right) , \;\; 
%%%
E^\mu_{(Z)} = 
\frac{1}{\sqrt{g_{\theta\theta}}}
\left(
\begin{array}{c}
0 \\
0 \\
1 \\ 
0 
\end{array}
\right) ,
\ee
with inverse
\be\label{eq-tetrad-inverse-lnro}
E^{(T)}_\mu = \sqrt{\frac{ g_{t\phi}^2 - g_{tt} g_{\phi\phi}}{g_{\phi\phi}}}
\left(
\begin{array}{c}
1 \\
0 \\
0 \\ 
0
\end{array}
\right) , \;\;
%%%
E^{(X)}_\mu = 
\sqrt{g_{rr}}
\left(
\begin{array}{c}
0 \\
1 \\
0 \\ 
0 
\end{array}
\right) , \;\;
%%%
E^{(Y)}_\mu = 
\sqrt{g_{\phi\phi}}
\left(
\begin{array}{c}
- \omega \\
0 \\
0 \\ 
1
\end{array}
\right) , \;\; 
%%%
E^{(Z)}_\mu = 
\sqrt{g_{\theta\theta}}
\left(
\begin{array}{c}
0 \\
0 \\
1 \\ 
0 
\end{array}
\right) .
\ee

%%%%%%%%%%%%%%%%%%%%%%%%%%%%%%%

\section{Proper areas of the annuli of an accretion disk}\label{a2}

Let us consider an observer in an equatorial circular orbit with Keplerian angular velocity. The tetrad and the inverse of the tetrad associated to such an observer are given in Eq.~(\ref{eq-tetrad}) and Eq.~(\ref{eq-tetrad-inverse}) with $\omega = \Omega_{\rm K}$, where $\Omega_{\rm K}$ is the angular velocity in Eq.~(\ref{eq-omega}).

First, we want to figure out the proper length of the orbit of this observer. At a certain time $t = t_*$, in the coordinate system $(t,r,\theta,\phi)$ we measure the angle $d\phi$. In his/her locally Minkowskian reference frame, the observer measures 
\be
dY = E^{(Y)}_\mu dx^\mu 
= u^t \sqrt{g^2_{t\phi} - g_{tt} g_{\phi\phi} } \, d\phi 
= \sqrt{\frac{g^2_{t\phi} - g_{tt} g_{\phi\phi}}{-g_{tt} - 2 g_{t\phi} \Omega_{\rm K} - g_{\phi\phi} \Omega_{\rm K}^2}} \, d\phi \, , 
\ee
because $dt = 0$. If the radial coordinate of the orbit is $r = r_i$, the proper length of the orbit of the observer is
\be
\ell (r_i) =
\int_0^{2\pi} d\phi \, \sqrt{\frac{g^2_{t\phi} - g_{tt} g_{\phi\phi}}{-g_{tt} - 2 g_{t\phi} \Omega_{\rm K} - g_{\phi\phi} \Omega_{\rm K}^2}}
= 2 \pi \left[ \sqrt{\frac{g^2_{t\phi} - g_{tt} g_{\phi\phi}}{-g_{tt} - 2 g_{t\phi} \Omega_{\rm K} - g_{\phi\phi} \Omega_{\rm K}^2}} \right]_{r = r_i , \theta = \pi/2} \, . 
\ee
Second, we want to figure out how the same observer measures distances perpendicular to his/her trajectory. If in the coordinate system $(t,r,\theta,\phi)$ we measure $dr$, the observer measures 
\be
dX = \sqrt{g_{rr}} dr \, .
\ee
We can now combine these two results to evaluate the proper area of an annulus of a Keplerian accretion disk with radial coordinate $r_i$ and width $\Delta r_i$:  
\be\label{eq-area}
A (r_i , \Delta r_i) = \int_0^{2\pi} d\phi \int_{r_i}^{r_i + \Delta r_i} dr \, \sqrt{g_{rr}}
\sqrt{\frac{g^2_{t\phi} - g_{tt} g_{\phi\phi}}{-g_{tt} - 2 g_{t\phi} \Omega_{\rm K} - g_{\phi\phi} \Omega_{\rm K}^2}}
= 2\pi \Delta r_i
\left[ \sqrt{\frac{g_{rr} \left( g^2_{t\phi} - g_{tt} g_{\phi\phi} \right)}{-g_{tt} - 2 g_{t\phi} \Omega_{\rm K} - g_{\phi\phi} \Omega_{\rm K}^2}} \right]_{r = r_i, \theta = \pi/2}
 \, .
\ee
This result is used in Eq.~(\ref{eq-properarea}).

In the existing literature, the calculation of $A (r_i , \Delta r_i)$ is normally proposed with a slightly different approach. First, we consider a locally non-rotating observer on the equatorial plane. The angular velocity of such an observer is $\omega = - g_{t\phi}/g_{\phi\phi}$ and the tetrad and the inverse of the tetrad are given in Eq.~(\ref{eq-tetrad-lnro}) and Eq.~(\ref{eq-tetrad-inverse-lnro}), respectively. If in the coordinate system $(t,r,\theta,\phi)$ we measure the angle $d\phi$, in his/her locally Minkowskian reference frame, the locally non-rotating observer measures
\be
dY^{\rm LNRF} = E^{(Y)}_\mu dx^\mu 
= \sqrt{ g_{\phi\phi} } \, d\phi \, .
\ee
The proper length of the orbit of the locally non-rotating observer with radial coordinate $r = r_i$ is
\be
\ell^{\rm LNRF} = \int_0^{2\pi} d\phi \, \sqrt{g_{\phi\phi}} = 2 \pi \left[ \sqrt{g_{\phi\phi}} \right]_{r = r_i , \theta = \pi/2}
\ee
Concerning length measurements perpendicular to his/her trajectory, if in the coordinate system $(t,r,\theta,\phi)$ we measure $dr$, the locally non-rotating observer measures
\be
dX^{\rm LNRF} = \sqrt{g_{rr}} \, dr \, .
\ee
If we consider a Keplerian accretion disk, the velocity of the material of the disk with respect to the locally non-rotating observer is $u^{(\alpha)} = E^{(\alpha)}_\mu u^\mu$, where $E^{(\alpha)}_\mu$ is given in Eq.~(\ref{eq-tetrad-inverse-lnro}), $u^\mu = u^t (1 , 0 , 0 , \Omega_{\rm K})$ is the 4-velocity of the material in the coordinate system $(t,r,\theta,\phi)$, and $u^t$ is given in Eq.~(\ref{eq-tdot2}). Since in a Minkowskian (or locally Minkowskian) reference frame the temporal component of the 4-velocity of a particle corresponds to its Lorentz factor~\cite{Bambi:2018drb}, we can evaluate the Lorentz factor $\gamma$ of the material of the accretion disk measured by the locally non-rotating observer as  
\be\label{eq-a2-gamma}
\gamma = u^{(T)} = E^{(T)}_\mu u^\mu 
= \left[ \sqrt{\frac{g_{t\phi}^2 - g_{tt} g_{\phi\phi}}{g_{\phi\phi} \left( -g_{tt} - 2 g_{t\phi} \Omega_{\rm K} - g_{\phi\phi} \Omega_{\rm K}^2 \right) }  } 
\right]_{r = r_i, \theta = \pi/2} \, .
\ee
The proper area of an annulus of a Keplerian accretion disk with radial coordinate $r_i$ and width $\Delta r_i$ can be evaluated as
\be\label{eq-area-method2}
A (r_i , \Delta r_i) = \gamma A^{\rm LNRF} (r_i , \Delta r_i)  \, ,
\ee
where $A^{\rm LNRF} (r_i , \Delta r_i)$ is the proper area of the annulus with radial coordinate $r_i$ and width $\Delta r_i$ in the rest-frame of the locally non-rotating observer 
\be
A^{\rm LNRF} (r_i , \Delta r_i) = \int_0^{2\pi} d\phi \int_{r_i}^{r_i + \Delta r_i} dr \, \sqrt{g_{rr} g_{\phi\phi}}
= 2 \pi \left[ \sqrt{g_{rr} g_{\phi\phi}} \right]_{r = r_i, \theta = \pi/2}\, \Delta r_i \, .
\ee
Eq.~(\ref{eq-area-method2}) is equivalent to Eq.~(\ref{eq-area}), of course.

%%%%%%%%%%%%%%%%%%%%%%%%%%%%%%%

\end{document}